\def\aatt{@}

%
\newcount\driver \newcount\mgnf \newcount\tipi
\newskip\ttglue
\def\TIPITOT{
\font\dodicirm=cmr12
\font\dodicii=cmmi12
\font\dodicisy=cmsy10 scaled\magstep1
\font\dodiciex=cmex10 scaled\magstep1
\font\dodiciit=cmti12
\font\dodicitt=cmtt12
\font\dodicibf=cmbx12 scaled\magstep1
\font\dodicisl=cmsl12
\font\ninerm=cmr9
\font\ninesy=cmsy9
\font\eightrm=cmr8
\font\eighti=cmmi8
\font\eightsy=cmsy8
\font\eightbf=cmbx8
\font\eighttt=cmtt8
\font\eightsl=cmsl8
\font\eightit=cmti8
\font\seirm=cmr6
\font\seibf=cmbx6
\font\seii=cmmi6
\font\seisy=cmsy6
\font\dodicitruecmr=cmr10 scaled\magstep1
\font\dodicitruecmsy=cmsy10 scaled\magstep1
\font\tentruecmr=cmr10
\font\tentruecmsy=cmsy10
\font\eighttruecmr=cmr8
\font\eighttruecmsy=cmsy8
\font\seventruecmr=cmr7
\font\seventruecmsy=cmsy7
\font\seitruecmr=cmr6
\font\seitruecmsy=cmsy6
\font\fivetruecmr=cmr5
\font\fivetruecmsy=cmsy5
\textfont\truecmr=\tentruecmr
\scriptfont\truecmr=\seventruecmr
\scriptscriptfont\truecmr=\fivetruecmr
\textfont\truecmsy=\tentruecmsy
\scriptfont\truecmsy=\seventruecmsy
\scriptscriptfont\truecmr=\fivetruecmr
\scriptscriptfont\truecmsy=\fivetruecmsy
\def \ottopunti{\def\rm{\fam0\eightrm}
\textfont0=\eightrm \scriptfont0=\seirm \scriptscriptfont0=\fiverm
\textfont1=\eighti \scriptfont1=\seii   \scriptscriptfont1=\fivei
\textfont2=\eightsy \scriptfont2=\seisy   \scriptscriptfont2=\fivesy
\textfont3=\tenex \scriptfont3=\tenex   \scriptscriptfont3=\tenex
\textfont\itfam=\eightit  \def\it{\fam\itfam\eightit}%
\textfont\slfam=\eightsl  \def\sl{\fam\slfam\eightsl}%
\textfont\ttfam=\eighttt  \def\tt{\fam\ttfam\eighttt}%
\textfont\bffam=\eightbf  \scriptfont\bffam=\seibf
\scriptscriptfont\bffam=\fivebf  \def\bf{\fam\bffam\eightbf}%
\tt \ttglue=.5em plus.25em minus.15em
\setbox\strutbox=\hbox{\vrule height7pt depth2pt width0pt}%
\normalbaselineskip=9pt
\let\sc=\seirm  \let\big=\eightbig  \normalbaselines\rm
\textfont\truecmr=\eighttruecmr
\scriptfont\truecmr=\seitruecmr
\scriptscriptfont\truecmr=\fivetruecmr
\textfont\truecmsy=\eighttruecmsy
\scriptfont\truecmsy=\seitruecmsy
}\let\nota=\ottopunti}
\newfam\msbfam   
\newfam\truecmr  
\newfam\truecmsy 
\def\TIPI{
\font\eightrm=cmr8
\font\eighti=cmmi8
\font\eightsy=cmsy8
\font\eightbf=cmbx8
\font\eighttt=cmtt8
\font\eightsl=cmsl8
\font\eightit=cmti8
\font\tentruecmr=cmr10
\font\tentruecmsy=cmsy10
\font\eighttruecmr=cmr8
\font\eighttruecmsy=cmsy8
\font\seitruecmr=cmr6
\textfont\truecmr=\tentruecmr
\textfont\truecmsy=\tentruecmsy
\def \ottopunti{\def\rm{\fam0\eightrm}
\textfont0=\eightrm
\textfont1=\eighti
\textfont2=\eightsy
\textfont3=\tenex \scriptfont3=\tenex   \scriptscriptfont3=\tenex
\textfont\itfam=\eightit  \def\it{\fam\itfam\eightit}%
\textfont\slfam=\eightsl  \def\sl{\fam\slfam\eightsl}%
\textfont\ttfam=\eighttt  \def\tt{\fam\ttfam\eighttt}%
\textfont\bffam=\eightbf
\def\bf{\fam\bffam\eightbf}%
\tt \ttglue=.5em plus.25em minus.15em
\setbox\strutbox=\hbox{\vrule height7pt depth2pt width0pt}%
\normalbaselineskip=9pt
\let\sc=\seirm  \let\big=\eightbig  \normalbaselines\rm
\textfont\truecmr=\eighttruecmr
\scriptfont\truecmr=\seitruecmr
}\let\nota=\ottopunti}
\def\TIPIO{
\font\setterm=amr7 
\font\settesy=amsy7 \font\settebf=ambx7 
\def \settepunti{\def\rm{\fam0\setterm}
\textfont0=\setterm   
\textfont2=\settesy   
\textfont\bffam=\settebf  \def\bf{\fam\bffam\settebf}
\normalbaselineskip=9pt\normalbaselines\rm
}\let\nota=\settepunti}

%
%

\newdimen\xshift \newdimen\xwidth \newdimen\yshift
%
%
\def\ins#1#2#3{\vbox to0pt{\kern-#2 \hbox{\kern#1 #3}\vss}\nointerlineskip}
%
%
%
\def\insertplot#1#2#3#4#5{\par%
\xwidth=#1 \xshift=\hsize \advance\xshift by-\xwidth \divide\xshift by 2%
\yshift=#2 \divide\yshift by 2%
\line{\hskip\xshift \vbox to #2{\vfil%
\ifnum\driver=0 #3
\special{ps: plotfile #4.ps} 
\fi
\ifnum\driver=1 #3 \includegraphics{#4.ps}\fi
\ifnum\driver=2 #3 \special{
\ifnum\mgnf=0 #4.ps 1. 1. scale \fi
\ifnum\mgnf=1 #4.ps 1.2 1.2 scale\fi} \special{ini.ps}
\fi }\hfill \raise\yshift\hbox{#5}}}


\let\a=\alpha \let\b=\beta   \let\d=\delta  \let\e=\varepsilon
 \let\g=\gamma

     \let\L=\Lambda 
      \let\X=\Xi
\let\Y=\Upsilon
%
%
\def\data{\number\day/\ifcase\month\or gennaio \or febbraio \or marzo \or
aprile \or maggio \or giugno \or luglio \or agosto \or settembre
\or ottobre \or novembre \or dicembre \fi/\number\year;\,\the\time}
\newcount\pgn \pgn=1
\def\foglio{\number\numsec:\number\pgn
\global\advance\pgn by 1}
\def\foglioa{A\number\numsec:\number\pgn
\global\advance\pgn by 1}
%

%
%
\global\newcount\numsec
\global\newcount\numfor
\global\newcount\numtheo
\global\advance\numtheo by 1

\def\senondefinito#1{\expandafter\ifx\csname#1\endcsname\relax}
\def\SIA #1,#2,#3 {\senondefinito{#1#2}%
\expandafter\xdef\csname #1#2\endcsname{#3}\else
\write16{???? ma #1,#2 e' gia' stato definito !!!!} \fi}
\def\etichetta(#1){(\veroparagrafo.\veraformula)%
\SIA e,#1,(\veroparagrafo.\veraformula) %
\global\advance\numfor by 1%
\write15{\string\FU (#1){\equ(#1)}}%
\write16{ EQ #1 ==> \equ(#1) }}
\def\letichetta(#1){\veroparagrafo.\verotheo
\SIA e,#1,{\veroparagrafo.\verotheo}
\global\advance\numtheo by 1
\write15{\string\FU (#1){\equ(#1)}}
\write16{ Sta \equ(#1) == #1 }}
\def\letichettaa(#1){A.\verotheo
\SIA e,#1,{A.\verotheo}
\global\advance\numtheo by 1
\write15{\string\FU (#1){\equ(#1)}}
\write16{ Sta \equ(#1) == #1 }}
\def\tetichetta(#1){\veroparagrafo.\veraformula 
\SIA e,#1,{(\veroparagrafo.\veraformula)}
\global\advance\numfor by 1
\write15{\string\FU (#1){\equ(#1)}}
\write16{ tag #1 ==> \equ(#1)}}
\def\FU(#1)#2{\SIA fu,#1,#2 }
\def\etichettaa(#1){(A.\veraformula)%
\SIA e,#1,(A.\veraformula) %
\global\advance\numfor by 1%
\write15{\string\FU (#1){\equ(#1)}}%
\write16{ EQ #1 ==> \equ(#1) }}
\def\alato(#1){}
\def\aolado(#1){}
\def\veroparagrafo{\number\numsec}
\def\veraformula{\number\numfor}
\def\verotheo{\number\numtheo}

\def\Eq(#1){\eqno{\etichetta(#1)\alato(#1)}}
\def\eq(#1){\etichetta(#1)\alato(#1)}
\def\leq(#1){\leqno{\aolado(#1)\etichetta(#1)}}
\def\teq(#1){\tag{\aolado(#1)\tetichetta(#1)\alato(#1)}}
\def\Eqa(#1){\eqno{\etichettaa(#1)\alato(#1)}}
\def\eqa(#1){\etichettaa(#1)\alato(#1)}
\def\eqv(#1){\senondefinito{fu#1}$\clubsuit$#1
\write16{#1 non e' (ancora) definito}%
\else\csname fu#1\endcsname\fi}
\def\equ(#1){\senondefinito{e#1}\eqv(#1)\else\csname e#1\endcsname\fi}
%
\def\Lemma(#1){\aolado(#1)Lemma \letichetta(#1)}%
\def\Lemmaa(#1){\aolado(#1)Lemma \letichettaa(#1)}%
\def\Theorem(#1){{\aolado(#1)Theorem \letichetta(#1)}}%
\def\Proposition(#1){\aolado(#1){Proposition \letichetta(#1)}}%
\def\Corollary(#1){{\aolado(#1)Corollary \letichetta(#1)}}%
\def\Remark(#1){{\noindent\aolado(#1){\bf Remark \letichetta(#1).}}}%
\def\Definition(#1){{\noindent\aolado(#1){\bf Definition
\letichetta(#1)$\!\!$\hskip-1.6truemm}}}
\def\Example(#1){\aolado(#1) Example \letichetta(#1)$\!\!$\hskip-1.6truemm}

\def\include#1{
\openin13=#1.aux \ifeof13 \relax \else
\input #1.aux \closein13 \fi}
\openin14=\jobname.aux \ifeof14 \relax \else
\input \jobname.aux \closein14 \fi
\openout15=\jobname.aux
%

%
%

\def\dd{{\rm d}}
 
 \def\\{\noindent}

 \def\V#1{\vec#1}

\def\tende#1{\vtop{\ialign{##\crcr\rightarrowfill\crcr
              \noalign{\kern-1pt\nointerlineskip}
              \hskip3.pt${\scriptstyle #1}$\hskip3.pt\crcr}}}
\def\otto{{\kern-1.truept\leftarrow\kern-5.truept\to\kern-1.truept}}
 \def\Z{{\bf Z}}\def\R{{\bf R}}
\def\mbox{\hbox}

  \def\LL{{\cal L}}

\def\={{\equiv}}
\def\initfiat#1#2#3{
\mgnf=#1
\driver=#2
\tipi=#3
\ifnum\tipi=0\TIPIO \else\ifnum\tipi=1 \TIPI\else \TIPITOT\fi\fi
%
%
\ifnum\mgnf=0
\magnification=\magstep0 \hoffset=0.cm
\voffset=-1truecm\hoffset=-.5truecm\hsize=16.5truecm \vsize=25.truecm
\baselineskip=14pt  
\parindent=12pt
\lineskip=4pt\lineskiplimit=0.1pt      \parskip=0.1pt plus1pt
\def\ds{\displaystyle}\def\st{\scriptstyle}\def\sst{\scriptscriptstyle}
\font\seven=cmr7
\fi
\ifnum\mgnf=1
\magnification=\magstep1
\hoffset=0.cm
\voffset=-1truecm
\hoffset=-.5truecm
\hsize=16.5truecm
\vsize=25truecm
\baselineskip=12pt
\parindent=12pt
\lineskip=4pt\lineskiplimit=0.1pt\parskip=0.1pt plus1pt
\def\ds{\displaystyle}\def\st{\scriptstyle}\def\sst{\scriptscriptstyle}
\font\seven=cmr7
\fi
\setbox200\hbox{$\scriptscriptstyle \data $}
}
\initfiat {1}{1}{2}

%
\expandafter\ifx\csname amssym.def\endcsname\relax \else\endinput\fi
%
\expandafter\edef\csname amssym.def\endcsname{%
       \catcode`\noexpand\@=\the\catcode`\@\space}
\catcode`\@=11
%
%
\def\undefine#1{\let#1\undefined}
\def\newsymbol#1#2#3#4#5{\let\next@\relax
 \ifnum#2=\@ne\let\next@\msafam@\else
 \ifnum#2=\tw@\let\next@\msbfam@\fi\fi
 \mathchardef#1="#3\next@#4#5}
\def\mathhexbox@#1#2#3{\relax
 \ifmmode\mathpalette{}{\m@th\mathchar"#1#2#3}%
 \else\leavevmode\hbox{$\m@th\mathchar"#1#2#3$}\fi}
\def\hexnumber@#1{\ifcase#1 0\or 1\or 2\or 3\or 4\or 5\or 6\or 7\or 8\or
 9\or A\or B\or C\or D\or E\or F\fi}
\font\tenmsa=msam10
\font\sevenmsa=msam7
\font\fivemsa=msam5
\newfam\msafam
\textfont\msafam=\tenmsa
\scriptfont\msafam=\sevenmsa
\scriptscriptfont\msafam=\fivemsa
\edef\msafam@{\hexnumber@\msafam}
\mathchardef\dabar@"0\msafam@39
\def\dashrightarrow{\mathrel{\dabar@\dabar@\mathchar"0\msafam@4B}}
\def\dashleftarrow{\mathrel{\mathchar"0\msafam@4C\dabar@\dabar@}}

\def\ulcorner{\delimiter"4\msafam@70\msafam@70 }
\def\urcorner{\delimiter"5\msafam@71\msafam@71 }
\def\llcorner{\delimiter"4\msafam@78\msafam@78 }
\def\lrcorner{\delimiter"5\msafam@79\msafam@79 }
\def\yen{{\mathhexbox@\msafam@55 }}
\def\checkmark{{\mathhexbox@\msafam@58 }}
\def\circledR{{\mathhexbox@\msafam@72 }}
\def\maltese{{\mathhexbox@\msafam@7A }}
\font\tenmsb=msbm10
\font\sevenmsb=msbm7
\font\fivemsb=msbm5
\newfam\msbfam
\textfont\msbfam=\tenmsb
\scriptfont\msbfam=\sevenmsb
\scriptscriptfont\msbfam=\fivemsb
\edef\msbfam@{\hexnumber@\msbfam}
\def\Bbb#1{{\fam\msbfam\relax#1}}
\def\widehat#1{\setbox\z@\hbox{$\m@th#1$}%
 \ifdim\wd\z@>\tw@ em\mathaccent"0\msbfam@5B{#1}%
 \else\mathaccent"0362{#1}\fi}
\def\widetilde#1{\setbox\z@\hbox{$\m@th#1$}%
 \ifdim\wd\z@>\tw@ em\mathaccent"0\msbfam@5D{#1}%
 \else\mathaccent"0365{#1}\fi}
\font\teneufm=eufm10
\font\seveneufm=eufm7
\font\fiveeufm=eufm5
\newfam\eufmfam
\textfont\eufmfam=\teneufm
\scriptfont\eufmfam=\seveneufm
\scriptscriptfont\eufmfam=\fiveeufm

%
\csname amssym.def\endcsname
%
%
%
%
\def\sqr#1#2{{\vcenter{\vbox{\hrule height.#2pt
     \hbox{\vrule width.#2pt height#1pt \kern#1pt
   \vrule width.#2pt}\hrule height.#2pt}}}}
\def\qed{ $\mathchoice\sqr64\sqr64\sqr{2.1}3\sqr{1.5}3$}
\def\IZ{Z\!\!\!Z\,}

\def\11{\hbox{l}\!\!\!1\,}

\font\tenib=cmmib10
\newfam\mitbfam
\textfont\mitbfam=\tenib
\scriptfont\mitbfam=\seveni
\scriptscriptfont\mitbfam=\fivei



\def\and{ \hbox{ and } }

\def\L{\Lambda}
\def\e{\varepsilon}
\def\a{\alpha}

\def\d{\delta}
\def\g{\gamma}

\def\to{\rightarrow}

\def\frac{\over}
\def\\{\cr}
\def\ref{}

\hbox{}
\vfill
\baselineskip12pt
\overfullrule=0in

\def\a{\alpha}

\def\d{\delta}

\def\text{\hbox}
\def\\\ {\cr}

%
%
\catcode`\X=12\catcode`\@=11
\def\n@wcount{\alloc@0\count\countdef\insc@unt}
\def\n@wwrite{\alloc@7\write\chardef\sixt@@n}
\def\n@wread{\alloc@6\read\chardef\sixt@@n}
\def\crossrefs#1{\ifx\alltgs#1\let\tr@ce=\alltgs\else\def\tr@ce{#1,}\fi
   \n@wwrite\cit@tionsout\openout\cit@tionsout=\jobname.cit
   \write\cit@tionsout{\tr@ce}\expandafter\setfl@gs\tr@ce,}
\def\setfl@gs#1,{\def\@{#1}\ifx\@\empty\let\next=\relax
   \else\let\next=\setfl@gs\expandafter\xdef
   \csname#1tr@cetrue\endcsname{}\fi\next}
\newcount\sectno\sectno=0\newcount\subsectno\subsectno=0\def\r@s@t{\relax}
\def\resetall{\global\advance\sectno by 1\subsectno=0
  \gdef\firstpart{\number\sectno}\r@s@t}
\def\resetsub{\global\advance\subsectno by 1
   \gdef\firstpart{\number\sectno.\number\subsectno}\r@s@t}
\def\v@idline{\par}\def\firstpart{\number\sectno}
\def\l@c@l#1X{\firstpart.#1}\def\gl@b@l#1X{#1}\def\t@d@l#1X{{}}
\def\m@ketag#1#2{\expandafter\n@wcount\csname#2tagno\endcsname
     \csname#2tagno\endcsname=0\let\tail=\alltgs\xdef\alltgs{\tail#2,}%
  \ifx#1\l@c@l\let\tail=\r@s@t\xdef\r@s@t{\csname#2tagno\endcsname=0\tail}\fi
   \expandafter\gdef\csname#2cite\endcsname##1{\expandafter
     \ifx\csname#2tag##1\endcsname\relax?\else{\rm\csname#2tag##1\endcsname}\fi
    \expandafter\ifx\csname#2tr@cetrue\endcsname\relax\else
     \write\cit@tionsout{#2tag ##1 cited on page \folio.}\fi}%
   \expandafter\gdef\csname#2page\endcsname##1{\expandafter
     \ifx\csname#2page##1\endcsname\relax?\else\csname#2page##1\endcsname\fi
     \expandafter\ifx\csname#2tr@cetrue\endcsname\relax\else
     \write\cit@tionsout{#2tag ##1 cited on page \folio.}\fi}%
   \expandafter\gdef\csname#2tag\endcsname##1{\global\advance
     \csname#2tagno\endcsname by 1%
   \expandafter\ifx\csname#2check##1\endcsname\relax\else%
\fi
   \expandafter\xdef\csname#2check##1\endcsname{}%
   \expandafter\xdef\csname#2tag##1\endcsname
     {#1\number\csname#2tagno\endcsnameX}%
   \write\t@gsout{#2tag ##1 assigned number \csname#2tag##1\endcsname\space
      on page \number\count0.}%
   \csname#2tag##1\endcsname}}%
\def\m@kecs #1tag #2 assigned number #3 on page #4.%
   {\expandafter\gdef\csname#1tag#2\endcsname{#3}
   \expandafter\gdef\csname#1page#2\endcsname{#4}}
\def\re@der{\ifeof\t@gsin\let\next=\relax\else
    \read\t@gsin to\t@gline\ifx\t@gline\v@idline\else
    \expandafter\m@kecs \t@gline\fi\let \next=\re@der\fi\next}
\def\t@gs#1{\def\alltgs{}\m@ketag#1e\m@ketag#1s\m@ketag\t@d@l p
    \m@ketag\gl@b@l r \n@wread\t@gsin\openin\t@gsin=\jobname.tgs \re@der
    \closein\t@gsin\n@wwrite\t@gsout\openout\t@gsout=\jobname.tgs }
\outer\def\localtags{\t@gs\l@c@l}
\outer\def\globaltags{\t@gs\gl@b@l}
\outer\def\newlocaltag#1{\m@ketag\l@c@l{#1}}
\outer\def\newglobaltag#1{\m@ketag\gl@b@l{#1}}

\def\t@gsoff#1,{\def\@{#1}\ifx\@\empty\let\next=\relax\else\let\next=\t@gsoff
   \expandafter\gdef\csname#1cite\endcsname{\relax}
   \expandafter\gdef\csname#1page\endcsname##1{?}
   \expandafter\gdef\csname#1tag\endcsname{\relax}\fi\next}
\def\verbatimtags{\let\ift@gs=\iffalse\ifx\alltgs\relax\else
   \expandafter\t@gsoff\alltgs,\fi}
\catcode`\X=11 \catcode`\@=\active
\localtags
%
 \let\epsilon=\varepsilon 
  



\let\epsilon=\varepsilon
\def\ds{\displaystyle}
\def\st{\scriptstyle}
\def\sst{\scriptscriptstyle}

\def\1{\,\rlap{\ninerm 1}\kern.15em 1}

\def\cite#1{[#1]}

\def\text#1{{\rm #1}}

\magnification=\magstep1
\hoffset=0.cm
\voffset=-.2truecm
\hoffset=-.2truecm
\hsize=16.5truecm
\vsize=24truecm
\baselineskip=14pt
\parindent=12pt
\lineskip=4pt\lineskiplimit=0.1pt\parskip=0.1pt plus1pt
\magnification=\magstep1
\hoffset=0.cm
\voffset=-.2truecm
\hoffset=-.2truecm
\hsize=16.5truecm
\vsize=24truecm
\parindent=12pt \lineskip=4pt\lineskiplimit=0.1pt
\parskip=0.1pt plus1pt
\def\ds{\displaystyle}\def\st{\scriptstyle}\def\sst{\scriptscriptstyle}

\def\picture #1 by #2 (#3){
  \vbox to #2{
    \hrule width #1 height 0pt depth 0pt
    \vfill
    \special{picture #3} 
    }
  }

\def\scaledpicture #1 by #2 (#3 scaled #4){{
  \dimen0=#1 \dimen1=#2
  \divide\dimen0 by 1000 \multiply\dimen0 by #4
  \divide\dimen1 by 1000 \multiply\dimen1 by #4
  \picture \dimen0 by \dimen1 (#3 scaled #4)}
  }

\def\IR{{\rm I\kern -1.8pt{\rm R}}}
\def\X{{\rm X\kern -6.0pt{\rm X}}}
\def\LL{{\rm I\kern -1.8pt{\rm L}}}
\def\W{{\rm W\kern -9.5pt{\rm W}}}
\def\V{{\rm V\kern -6.5pt{\rm V}}}
\def\TT{{\rm I\kern -4.3pt{\rm T}}}
\def\U{{\rm U\kern -6.0pt{\rm U}}}
\def\Y{{\rm Y\kern -6.0pt{\rm Y}}}

\def\IE{{\rm I\kern -1.6pt{\rm E}}}

\def\IP{{\rm I\kern -1.6pt{\rm P}}}
\def\IZ{{\rm Z\kern -4.0pt{\rm Z}}}
\def\IC{\ {\rm I\kern -6.0pt{\rm C}}}
\def\TT{{\rm I\kern -4.3pt{\rm T}}}

\initfiat {1}{1}{2}

\def\b{\beta}

\def\leq{\le}
\def\NoBlackBoxes{\global\overfullrule0pt}
\NoBlackBoxes

\font\titlefont=cmbx10 at 14.4 truept

\magnification=1200


\centerline {\titlefont Macroscopic Evolution of Particle Systems}

\centerline {\titlefont with Short and Long Range Interactions }
\vskip 1truecm
\centerline{\baselineskip=10pt
Giambattista Giacomin
\footnote{$^1$}{\eightrm Dipartimento di Matematica, Universit\`a di Milano, via 
C. Saldini 50, 20133 MI, Italy. E--mail: giambattista.giacomin\aatt  mat.unimi.it},
\hskip.2cm
Joel L. Lebowitz \footnote{$^2$}
{\eightrm  Departments of Mathematics and
Physics, Hill Center. 110 Frelinghuysen Rd,
 Rutgers University, Piscataway, NJ 088954--8019, USA. E--mail: lebowitz\aatt
 math.rutgers.edu }
\hskip.2cm
and
\hskip.2cm
Rossana Marra
\footnote{$^3$}{\eightrm Dipartimento di Fisica and Unit\`a INFM, 
Universit\`a di Roma Tor Vergata, Via della Ricerca Scientifica,
00133 Roma, Italy. E--mail: rossana.marra\aatt roma2.infn.it} 
}

\bigskip
\bigskip
\medskip

\bigskip {\baselineskip = 10pt\rightskip1.4cm\leftskip 1.4cm \ottopunti
{\noindent {\bf Abstract:}\/} 
\quad 
We consider a lattice gas with general short range interactions and a Kac
potential $J_\gamma({r})$ of range $\gamma^{-1}$, $\gamma>0$, evolving via particles 
hopping to nearest neighbor  empty sites with rates  which satisfy detailed balance with
respect to the equilibrium measure. 
Scaling space like $\gamma^{-1}$ and
time like $\gamma^{-2}$,  we prove that  in the limit $\gamma \to 0$ the
macroscopic density profile $\rho({ r},t)$ satisfies the equation
$$
\phantom{recenterrecenter}
{\partial \over \partial t} \rho({ r},t) = \nabla \cdot \left[\sigma_s(\rho)
\nabla {\delta {\cal F} (\rho)\over \delta \rho({ r})}\right]. 
\ \ \ \ \ \ \ \ \ \ \ \  \  
\ \ \ \ \ \ \ \ \ \  \ \ (*)
$$
Here $\sigma_s(\rho)$ is the mobility of the reference system, the one with
$J \equiv 0$, and ${\cal F} (\rho)
= \int[f_s(\rho(r)) - {1 \over 2} \rho(r) \int
J({ r}-{ r}^\prime)\rho({ r}^\prime) {\rm d}r
{\rm d}{r}^\prime]$, where $f_s(\rho)$ is the (strictly convex) 
free energy density of the
reference system. 
Beside a regularity condition on $J$,
the only requirement for this result is  that 
the reference system satisfy the hypotheses of the Varadhan--Yau Theorem
[VY] leading to ($*$) for $J\equiv 0$. Therefore ($*$) holds also if  
 $\cal F$ achieves its minimum on non constant density profiles and
this includes the  cases in which {\sl phase segregation} occurs. Using 
the same techniques we also derive hydrodynamic equations for the 
 densities
of a two component A-B mixture with long range repulsive interactions
 between A and B particles. The equations for the densities $\rho_A$ and $\rho_B$
 are of the form ($*$). They describe, at low temperatures,
 the demixing transition in which
  segregation takes place via vacancies, i.e. jumps to empty sites.
In the limit of {\sl very few} vacancies 
the problem becomes similar to phase segregation in a continuum 
system in the so called {\sl incompressible limit}
[EP], [DG].

\bigskip
\bigskip}

\noindent
2000 MSC: 82C22, 60K35

\vskip 0.3 cm

\noindent
Key Words: {\it Interacting particle  
systems, Kac potential,
hydrodynamic 
limits, phase segregation, non--gradient 
particle models, Einstein relation, vacancy dynamics.}

\vskip.5cm

\numsec=1
\numfor=1

\vskip 0.4 cm

The state of a (one component) macroscopic system in equilibrium can be
characterized by two numbers, the temperature $T (= \beta^{-1})$ and the
chemical potential $\lambda$.  When $T$ and $\lambda$ correspond to a
single phase (i.e.\ there is a unique Gibbs measure) then the particle
density $\bar \rho(T, \lambda)$ is constant, i.e.\ spatially uniform. 
   Given a microscopic dynamics for which 
this Gibbs measure is attractive,  a disturbance in this uniform density
corresponding to a profile $\rho_0(r)$, $r$ the space coordinate, at some time $t_0$ 
is expected
to relax towards the uniform density $\bar \rho$.  In certain types of
systems (when the variations in temperature and hydrodynamical flows can be
neglected, see below) the  relaxation of the density profile $\rho(t,r)$ 
will occur via the diffusion equation
$$
{\partial \rho \over \partial t} = {\nabla} \cdot \left[D {
\nabla} \rho \right], \quad t > t_0, 
\Eq(1.1)
$$
where the bulk diffusion constant $D = D(\rho)$ will generally depend also
on the temperature $T$, assumed constant and therefore omitted
in the notation.

Equation \equ(1.1) is a continuity equation for $\rho$ corresponding to a mass
current given by Fick's law
$$
{ j} = -D(\rho) { \nabla} \rho. \Eq(1.2)
$$
To derive \equ(1.1) from microscopic models it is most convenient to write
Fick's law in its Onsagerian form,
$$
j = -\sigma \nabla \lambda , \Eq(1.3)
$$
where $\sigma$ is the {\sl conductivity}, or {\sl mobility}, and $\lambda(\rho)$ is the
local chemical potential; $\lambda$ and  $\rho$ are related as in the uniform equilibrium system, i.e. we are in a situation 
of {\sl local equilibrium}. 
 Comparing \equ(1.2)
and \equ(1.3) gives 
$$
D = \sigma/\chi \Eq(1.4),
$$
where
$$
\chi (\rho)  = \left({\partial  \lambda \over \partial \rho}\right)^{-1}=\rho{\partial\rho\over\partial p} \Eq(1.5chi)
$$
is the {\sl compressibility} ($p$ being the pressure).
In \equ(1.5chi) we have expressed the chemical potential at equilibrium
as a function of the density $\rho$.
 The relation \equ(1.4) is sometimes referred to as the {\sl Einstein relation} who
first used it to relate the diffusion constant of a Brownian particle  
to its steady state mobility in an external field 
[E, Sp].

Mathematically rigorous derivations of \equ(1.2) -- \equ(1.5chi), 
with $\sigma$ given by a Green--Kubo formula, have been achieved, via the use of the
hydrodynamical (diffusive) scaling limit for a variety of microscopic
models  with fixed short (microscopic) range interactions evolving via stochastic dynamics (in which particle number is the
only conserved quantity), see [KL]\footnote{$^{(1)}$}{There also exists a derivation for one
system with a Hamiltonian evolution, namely a non--interacting gas of point
particles moving among fixed periodic scatterers -- the Sinai billiard with
finite horizons [BS], [ Sp].}.  In all these cases the temperature $T$ is in
the uniqueness region of the phase diagram, i.e. there is a unique phase for all values 
of $\lambda$ (or $\rho$).

The situation becomes much trickier when we consider temperatures  
where there is, for some value of $\lambda$, a
 coexistence of phases with two (or more) different densities,
$\rho_1$ and $\rho_2$, corresponding to
 liquid and vapor or to fluid and solid.  In such cases
the macroscopic equilibrium system, with a fixed total  number of
particles corresponding to an average density in the 
interval $(\rho_1,\rho_2)$, 
 will not have a uniform density. Instead it will
be segregated into
macroscopic regions of density $\rho_1$ and $\rho_2$ with shapes minimizing
the free energy of the surface 
 between them.  An equation of form \equ(1.1) is
clearly not appropriate now, in fact, for any density $\rho \in (\rho_1,
\rho_2)$,  $\chi(\rho)$ in \equ(1.4) will be infinite and therefore
  $D$ is formally
zero (unless $\sigma$ is also
infinite, which can be proven in some cases not to happen).
Results in these directions have been proven for very particular
systems, see the remarkable result in [R] where degenerate
diffusion in the coexistence region is proven, but the
general  case of the evolution of phase domains
for systems with  interactions 
is a real challenge for the moment.

To get around this obstacle and derive macroscopic equations for a system
undergoing phase segregation, some authors  studied the time
evolution of the macroscopic density profile in particle systems
interacting via long range (compared to the interparticle spacing) Kac
potentials, [LOP], [GL1].  The microscopic model 
considered in [GL1] is a lattice gas evolving under a 
particles hopping (Kawasaki exchange) dynamics
which satisfies detailed balance (is reversible) with
respect to the Gibbs canonical (fixed particle number) equilibrium measure
with Hamiltonian $H$ at temperature $\beta^{-1}$.  $H$ consisted of a sum
of two terms, a short range part $H_s$ which may be thought of as a nearest
neighbor  interaction and a Kac potential
$H_\gamma$ characterized by a range parameter $\gamma^{-1}$, namely
$$
H_\gamma = -{1 \over 2} \sum_{x,y} \gamma^d J(\gamma|x-y|)\eta(x) \eta(y),
\Eq(1.5H)
$$
where $x,y \in \Lambda_\gamma \subset {\Bbb Z}^d$ (say a torus of diameter
$[\gamma^{-1}]$) and $\eta(x) = 0,1$
specifies the occupancy of site $x$. They argued that 
 in the diffusive hydrodynamical scaling limit, corresponding to $\gamma \to
0$ with space and
time scaling like $\gamma^{-1}$ and $\gamma^{-2}$ respectively, the density
profile $\rho(r,t)$ would satisfy a parabolic integro--differential 
equation of
the form,
$$
{\partial \rho \over \partial t} = \nabla \cdot\left[
\sigma_s(\rho) \nabla{\delta
{\cal F}(\rho) \over \delta \rho}\right]\equiv \nabla \cdot\left[
\sigma_s(\rho) \nabla \lambda(\rho)\right],
\Eq(1.6)
$$
where
$$
{\cal F} = \int f_s(\rho(r))dr - {1 \over 2} \int \int
J(r-r^\prime)\rho(r)\rho(r^\prime)dr dr^\prime\Eq(1.7)
$$
$f_s$ is the free energy density and $\sigma_s$ is the mobility of the
system with only short range interactions.

Equations \equ(1.5chi) and \equ(1.6) were proven in [GL1]
 for the case where the short range
interactions $H_s$ consisted only of the hard core exclusion, i.e. no more
than one particle per site, in which case $f_s$ and $\sigma_s$ reduce to
$f^0_s$ and $\sigma^0_s$, with
$$
f^0_s (\rho) = \beta[\rho \log \rho + (1 - \rho) \log(1 -
\rho)],\Eq(1.8)
$$
and
$$
\sigma^0_s(\rho) = \beta \rho(1 - \rho).\Eq(1.9)
$$

The validity of \equ(1.6) and \equ(1.7) for the case of nontrivial short range
interactions was conjectured in [\S 3,GL1] for values of $\beta^{-1}$ {\it above the
critical temperature} of the reference (short range) system,
 when $f_s(\rho)$ 
is a strictly convex function of $\rho$. They gave  heuristic
arguments for the validity of \equ(1.6) with $\lambda$ the {\sl local} chemical
potential of the system with the Kac potential, but with the mobility $\sigma$  being 
the same as what it would be in the system without  the slowly varying Kac
potential.  This  generalized a similar
conjecture by Spohn [Sp] for the case of an {\sl external long range
interaction}. Besides the case in [GL1], 
the conjecture was shown to be valid 
also for some other special cases
 [BL],[MM],[AX].    In this paper we prove
these conjectures for the general short range interactions in the case
of Ising spins, that is 
in the case of a system with Hamiltonian $H^\gamma=
H_s+H_\gamma$. In fact the methods we use here extend
 in a straightforward manner to general systems for which a diffusion
equation can be proven in the absence of $H_\gamma$. 
We illustrate this by deriving in Section 5 integro--differential
equations for
a binary mixture which may undergo a demixing transition.  Our results extend to the case of
a system on which a weak  external force is acting, characterized by the Hamiltonian
$$H_s(\eta) +\sum_x V(\gamma x)\eta(x)$$
and also to the case of a weak (of order $\gamma$) external force which is  not the gradient
of a potential, like a costant force  in a torus.

We note  that solutions to \equ(1.6) corresponding to interface dynamics 
are considered in
[GL2].  There  is  also a review of various  results about these models,
including the case in which there is no conservation law ({\sl
Glauber} dynamics) [GLP]. 
\vskip.1cm

{\it Informal Description of Model}.   To be  concrete, we restrict ourselves
to the case $H_s(\eta)= K \sum_{x,y:\vert x-y\vert=1}
\eta(x)\eta(y)$.   
 The  Kawasaki dynamics is defined in terms of 
Poisson jump rates depending on the energy differences. 
An example of such rates is
$$
c^\g_{(x,y)} (\eta)=\exp \left(
 -\beta \left[H^\g(\eta^{x,y})- H^\g(\eta)\right]
/2\right),
\Eq(0.5)
$$
where  $x, y$ are nearest neighbor sites and  $\eta^{x,y}$ is the configuration 
in which the sites $x$ and $y$ exchange
their occupation numbers.
The microscopic current $w_{x,y}$ through the bond $(x,y)$ is the rate at which a particle jumps from
$x$ to $y$ minus the rate at which a particle jumps from
$y$ to $x$, namely 
$$
w^\g_{x,y}(\eta)=\left[
\eta(x)c^\g_{(x,y)}(\eta)-\eta(y)c_{(x,y)}^\g(\eta)\right].
\Eq(5.3)
$$
Equation \equ(5.3) will 
determine the form of the macroscopic current in the hydrodynamic equation.

A key ingredient in our analysis is
 that the dynamics with $J\not=0$ is a {\sl weak} perturbation
of the $J\equiv 0$ dynamics. This can be seen both at the level of the rates
and of the current. In particular for the current some straightforward 
expansions with  respect to the small parameter $\gamma$ (see Section 4)
lead to 
$$
\eqalign{
w^\g_{x,x+e_i}(\eta)
=&c^0_{x,x+e_i}(\eta)\left[\eta(x)-\eta(x+e_i)\right]\cr
&+{\g\b \over 2} c^0 _{x, x+e_i} 
\left[ \eta(x)-\eta(x+e_i)\right]^2
 \left(
\partial_i J \star \eta\right) (x)
+ O(\gamma^{2}),
}
\Eq(0.6)
$$
where $e_i$ is the unit vector on the lattice in the direction $i$, 
 $\star$ denotes spatial discrete
convolution and  the superscript $0$ on $c$ and $w$  denotes the case in which
there is no long range force ($J\equiv 0$).
Equation \equ(0.6) indicates clearly that 
the dynamics with $J\neq 0$ is a weak perturbation of the
reference ($J\equiv 0$) dynamics. The case studied 
in [GL1] corresponds to $ c^0_{x,y}=1$
in \equ(0.6). The local equilibrium expectation of $w^0_{x,x+e_i}$ gives
the macroscopic flux $j$ for this simplified
model in  the form of a term due to the exclusion dynamics
and one due to the mean field force:
 $$
j=-\nabla\rho+ \sigma(\rho) \nabla J*\rho,
\Eq(mcfsm)
$$
 where $*$ denotes the (standard) spatial convolution, $J*\rho(r)=\int J(r-r^\prime) 
\rho (r^\prime) \dd r^\prime$,
and the mobility
$\sigma(\rho=
\beta\rho(1-\rho)$ is just 
 the expectation of $(\beta/2)\left[ \eta(x)-\eta(x+e_i)\right]^2$.
 
The naive extension of the above  argument to the general case, 
i.e. 
just applying local equilibrium to  \equ(0.6),
 would give  a macroscopic current of
the form
$$
j=-D\nabla\rho + \tilde\sigma (\rho) \nabla J*\rho,
\Eq(0.7)
$$
where $D$ is, as before, the diffusion coefficient
found in [VY], but  $\tilde \sigma$ would not satisfy the Einstein
relation \equ(1.4).
This apparent contradiction is explained by the fact that the 
microscopic current 
of the reference system in the general case, i.e. the first term in the
right--hand side of \equ(0.6),  is 
 in not  a lattice gradient of a function of the configuration,
i.e. there is no function $h$ such that $w^0_{x,y}(\eta)=
h(\tau_y \eta)-h (\tau_x \eta)$,
where $\tau_x$ is the translation in configuration space,  
unless $K=0$ (i.e. $H_s\equiv 0$).
 The non gradient nature of the dynamics is
responsible for the presence  in
\equ(0.7) of a third term which, combined with the second term, gives the correct expression for the
macroscopic current.

The main 
ingredient in our derivation is the recent work of Varadhan and Yau [VY]
who proved \equ(1.1) for a lattice gas with general short range  interactions
at  small $\beta$.  This is a major achievement since previous derivations
all required  that the dynamics be either of gradient type, 
or that the invariant measure be of product type, i.e.
independent occupation values at different sites, see [KL] for a complete
treatment and extended bibliography on this topic.
The class of non--gradient models with invariant measures of
product type includes the $n$--color simple exclusion process, which is a standard simple exclusion 
process but each  particle is colored with one of $n$ possible colors. 
 Exchanges only occur between occupied and empty sites, so that 
  two sites with different colors cannot exchange.
Our arguments do apply to this case too.
This can be viewed as  a multi--species system and the effect of 
long range interactions is of interest 
in its own right and we discuss it in Section 5 of this paper, limiting ourselves
to the $n=2$ case, corresponding to a binary alloy
in which exchanges take place only through vacancies.

The paper is organized as follows:
in section 2 we give  the precise definition of our model 
and we 
state the main results. In section 3 we recall 
some important results of [VY] that 
we need for our proof.
Section 4 contains the proofs which are based 
on the control of the Radon-Nykodim derivative of the full
process with respect to the reference process and on the [GPV]--[V] method. 
Finally in section 5 we consider the   
example, that we just mentioned, 
of a dynamics with two conservation laws, for which we prove a result analogous to
the one in section 2. Moreover, in this case we get  a stronger convergence result, since 
 uniqueness of the weak solution for the limit equation does hold, under suitable
hypotheses.

\vskip.8cm

\numsec=2
\numfor=1
\centerline{\bf 2. The model and the main result.}
\vskip 0.4 cm

We  work  in the discrete torus of dimension $d$ and 
diameter $\g^{-1}$, $\gamma>0$, that we denote by  
$\Lambda_\gamma$. 
 Associated to  $\Lambda_\g$ 
there is a natural notion of $\Lambda_\g^*$,
the set of (non directed) bonds, i.e. the couples
of nearest neighbor points of $\Lambda_\g$.  The configuration
space on $\Lambda_\g$ is
$\{0,1\}^{\Lambda_\g} \equiv \Omega_{\Lambda_\g}$. 
All these spaces are  endowed with the discrete topology:
when later we will deal with $\{0,1\}^\Lambda$,
$\Lambda$ countable, we will use the
product topology.

\vskip 0.2 cm
\noindent
{\bf 2.1 The Gibbsian reference measure.}
Let $F$ be a local isotropic function from $\Omega_{\IZ^d}$ to
$\IR$. By isotropic we mean that $F(\theta \eta)=
F(\eta)$ for every reflection $\theta$ with respect of an
axis as well any lattice rotation $\theta$.
For any $x\in \IZ^d$ set $F_x(\eta)=
F(\tau_x \eta)$, where $\tau_x$ is the translation
of $x\in \IZ^d$ in $\Omega_{\IZ^d}$:
$$
\tau_x \eta (y)= \eta (x+y),
\ \ \ \ \ \ \ \ {\rm for \ every \ } y \in 
\IZ^d. 
$$
And $\tau$ will also be the translation operator acting on functions
$F$ of the configuration: $\tau_x F(\eta)=F( \tau_x \eta)$.

Let us consider the formal Hamiltonian
$$
H(\eta) = \sum_x  F_x (\eta).
\Eq(Hamiltonian)
$$
It is well know that if the inverse temperature
 $\kappa$ is sufficiently small, given any 
chemical potential $\lambda$,
there 
exists a unique extremal  
Gibbs measure $\mu^\rho$,
$\rho \in [0,1]$, 
which satisfies $\rho=
\IE^{\mu^\rho}[\eta (x)]$ for every
$x \in \IZ^d$. We will always work in this
{\sl uniqueness} regime: this guaranties 
also that $\mu^\rho$ is translation invariant and that
it has some mixing properties. 
However in some proofs in [VY] the authors 
require a stronger mixing condition
({\sl exponential mixing condition}, [Assumption A, VY]), 
given in terms
of finite volume grand--canonical measures.
Since we are using this assumption only indirectly
we do not give it explicitly: we just stress that 
our results are proven only if $\kappa $ is smaller
then a certain $\kappa_0>0$, which depends on the 
dimension $ d$ and on the interaction $F$.
In some cases it can be shown that $\kappa_0= \kappa_c$,
the inverse of the critical temperature of the
reference system.

We will deal with the $F$--interaction
also in the case of $\Lambda_\gamma$:
for $\g$ sufficiently small
 we can keep the very same
infinite volume definitions by lifting
$\Lambda_\gamma$ to the whole of $\IZ^d$
in the natural way. This way the notion 
of translation $\tau_x$ is unchanged. It
is however often more natural to look
at the periodic case as a finite volume case,
and that's what we will do.
The same applies to every local function
on $\Omega_{{\bf Z}^d}$, which will be viewed
as a function on $\Omega_{\Lambda_\g}$
without notational changes.

\vskip 0.2 cm
\noindent
{\bf 2.2 The equilibrium measure for the full system.}
We consider a the probability measure on $\Omega_{\Lambda_\gamma}$
defined by 
$$
\mu_{\gamma}^{\beta, \kappa, \lambda}(\eta)=
{1 \over Z_\gamma (\beta, \kappa, \lambda) }
\exp\left\{
-\kappa \sum_x  F_x  - {\beta \over 2} \sum _{x,y}
J_\gamma (x,y) \eta (x) \eta (y) +
\lambda \sum_x \eta (x)
\right\}, 
\Eq(fullmeasure)
$$
$$
\beta>0, \gamma>0, \lambda \in \R,
$$
where
$J_\gamma(x,y)= \gamma ^d J(\gamma (x-y))$,
$J \in C^2 (\TT^d)
\rightarrow \IR$ 
($\TT^d$ is the d-dimensional torus of diameter $1$) such that 
$$
J(r)=J(-r),
\ \ \ \ \ \ \ \ \
\int_{\TT^d}
J(r) {\rm d}r =1,
\Eq(Jcond)
$$
and $Z_\gamma (\beta,\kappa,\lambda)$ is the normalization factor.

\vskip 0.2 cm
\noindent
{\bf 2.3 The dynamics.} This will be introduced, for the reference system, 
both in the case
$\Lambda=\Lambda_\g$ and  in the case $\Lambda=\Z^d$. 
Let $\Phi\in C^2 (\IR;\IR^+)$ 
be of the form\footnote{$^1$}{This is equivalent to 
the more customary [p. 163, Sp]
detailed balance condition $\Phi (E)=\Phi(-E) \exp(-E)$,
$E \in \IR$.}
$$
\eqalign{
& \Phi(E)=\exp(-E/2) \phi (E),
\cr
\phi:\IR \rightarrow \IR^+, \
&
\ \  
\phi(E)=\phi(-E) \ \ {\rm for \ every \ }
E\in \IR \ {\rm and }\ \phi(0)=1.
}
\Eq(detbalalt)
$$
For $b=(x,y) \in \Lambda^*$ we define 
$$
c_b^0 (\eta)=\Phi ( \kappa \Delta_b H (\eta)),
\Eq(prerates0)
$$
where $\Delta_b H( \eta) =
H(\eta^b)- H(\eta)$ and
$$
\eta ^b (z)=
\cases{
\eta (x), & if $z=y$, 
\cr
\eta (y), & if $z=x$, 
\cr
\eta (z), & otherwise.
}
\Eq(exchange)
$$
In the infinite volume case 
$H$ is not well defined, but $\Delta_b H$ is
taken, by definition, to be equal to $\lim_{R\rightarrow \infty}
\Delta_b \sum_{\vert x\vert \le R}
F_x (\eta)$.
To make the notation a bit lighter, 
if $b=(x,y)$ appears as a subscript, we will often 
drop the brackets.
Given $f:\Omega _\Lambda \rightarrow \IR$
for $b=(x,y)$ we set
$$
{\cal L }^0_b = c_b^0(\eta) 
\left[ 
f(\eta^b) -f(\eta)\right].
\Eq(L0b)
$$
A Markov pregenerator is then defined by setting
$$
{\cal L}^0 f(\eta)= \sum_{b\in \Lambda^*} {\cal L}^0_b f(\eta),
\Eq(pregen)
$$
where $f$ is assumed to be local in the case $\Lambda=
\IZ^d$.
If $\Lambda=\Lambda_\gamma$, ${\cal L}^0$ is actually
a generator and it is easy to construct a unique process
in Skorohod space
$\{ \eta_t \}_{t \in \IR^+}\in D(\IR^+;
\Omega_{\Lambda_\gamma})$ associated to it, once an
initial condition is given. The law of this process
will be denoted by ${\bf P}^{\g,0}$ or
${\bf P}^{\g, 0}_{\mu_\g}$ if there is the need
to stress the initial condition $\mu_\g$. 
It is immediate to verify that, by 
\equ(detbalalt),  ${\cal L}^0$ viewed as an
operator in
$L^2(\mu_{\gamma}^{0, \kappa, \lambda})$ is
self--adjoint for every $\lambda \in \IR$.
In the infinite volume setting the construction 
of the process is more delicate. We refer to  [Li]
for this construction: there the process 
$\{ \eta_t \}_{t \in \IR^+}\in D(\IR^+;
\Omega_{\IZ^d})$ associated to ${\cal L}^0$
is constructed.  We remark also that, for every
$ \rho \in  [0,1]$, ${\cal L}^0$ 
can be extended to a self--adjoint
operator on $L^2 (\mu^\rho)$. The law of the process
$\{ \eta _t\}_{t \in \IR^+}$ will be denoted by ${\bf P}^0$.
We will sometimes stress the chosen initial 
condition, say $\mu \in {\cal P}_1(\Omega _\Lambda)$, 
by writing ${\bf P}^0_\mu$. Here we used ${\cal P}_1 (\cdot)$
to denote the probability measures on $\cdot$.

The full dynamics is considered only in the case
$\Lambda=\Lambda_\gamma$. We define
$$
c_b^\gamma (\eta)=
\Phi\left( \Delta_b \left[
\kappa \sum_x  F_x  - \beta \sum _{x,y}
J_\gamma (x,y) \eta (x) \eta (y) \right]
\right),
\Eq(cbfull)
$$
and 
$$
{\cal L}_{\gamma} f (\eta)=
\sum_{b\in \Lambda^*}
 c_b^\gamma (\eta) \left[
f(\eta^b)-f(\eta) \right].
\Eq(fullgendyn)
$$
As before, associated to the finite dimensional
operator ${\cal L}_{\gamma}$ there is a process
with trajectories in $D(\IR^+; \Omega_{\Lambda_\gamma})$.
${\cal L}_{\gamma}$ is self--adjoint in
$L^2(\mu_{\gamma}^{\beta, \kappa, \lambda})$ for every 
$\lambda \in \IR^+$. The law of this process will be denoted 
by ${\bf P}^\gamma={\bf P}^\gamma_{\mu_\gamma}$,
where $\mu_\gamma \in {\cal P}( \Omega_{\Lambda_\gamma})$ 
is the initial condition.

\vskip 0.2 cm
\noindent
{\bf 2.4 The main result.}
The compressibility $\chi$ for the system 
 is defined
as
$$
\chi(\rho)=
\sum_{x\in \IZ^d} {\rm cov}^{\mu^\rho}\left(
\eta(0),\eta (x)
\right),
\Eq(compressibility)
$$
in terms of the local interaction alone.
The corresponding
diffusion matrix $D$ can be expressed via
a variational formula. It is the $\rho$ dependent
symmetric matrix defined by
$$
\langle v, Dv \rangle_{\IR^d}=
{1 \over 2\chi (\rho) }
\inf_{g} \IE^{\mu^\rho}\left[
\sum_{i=1}^d
c_{e_i}^0 (\eta)
\left(
v_i(\eta(e_i)-\eta(0))-
\sum_{x\in \IZ^d} \Delta_{(0,e_i)} g\left(\tau_x \eta \right)
\right)^2
\right],
\Eq(varform)
$$
for every $v \in \IR^d$.
In \equ(varform) $e_i$ denotes the unit vector in the
$i$ direction, $\langle \cdot, \cdot \rangle_{\IR^d}$
is the scalar product in $\IR^d$ and the infimum
is taken over all local functions $g$.
We will take the freedom of using both the notation
$v_i$ and the notation $v_{e_i}$. Moreover
for $e$ a unit vector, $v_e=\pm v_{e_i}$ if
$e=\pm e_i$.

A number of facts are known about $D$:  first of all it is
 a continuous function of $\rho$, cf. [VY],
and there exists a constant $c$ such that in the
sense of matrices
$$
{I \over c \chi(\rho)} \le D(\rho) \le cI,
\Eq(boundsonD)
$$
where $I$ is the $d\times d$ identity matrix.
While the upper bound is an immediate consequence of
\equ(varform), the lower bound is much more subtle
and it has been established in [SY].
In [VY, Lemma 8.3] it is shown moreover that in our
isotropic set up, $D(\rho)$ is a multiple of $I$.
We will keep the notation $D(\rho)$ also to
denote the scalar proportionality factor between
$I$ and the matrix $D(\rho)$.

We are going to consider initial particle configurations
associated to a density profile $\rho_0 : \TT \rightarrow [0,1]$,
in the following sense: if we define the 
 the empirical density field
$$
\nu^\gamma (t,x)=
\gamma^d \sum_{y\in\L} \delta (x-\gamma y) \eta_{\gamma^{-2}t}(y),
\Eq(empfield)
$$
we require that for any smooth test 
function $G$ from $\TT^d$ to $\IR$ and $\d>0$ 
$$
\lim_{\gamma\to 0} {\bf P}^{\gamma}_{\mu_\gamma}
\left(\left\vert \int_{\TT^d}\nu^\gamma (x,0)G(x)dx-\int
_{\TT^d}\rho_0(x)G(x){\rm d} x
\right\vert > \delta\right)=0.
\Eq(1.5)
$$

Denote  by
${\bf Q}^{{\g}}$ 
the law of the process $\{\nu^\g(t)\}_ {t\in [0,T]}$ on the space 
$D\big([0,T],{\cal M}\big)$, induced by ${\bf P}^{\g}_{\mu_\g}$, 
where ${\cal M}={\cal M}(\TT^d)$ is the space of nonnegative
measures on the torus with total mass bounded by $1$: this 
space is endowed with the weak topology weakened by the
continuous functions. 
We consider also the subspace ${\cal M}_1={\cal M}_1(\TT^d)$
of $\cal M$ consisting of absolutely continuous measures 
with density bounded  above by one. 
Our notation for the gradient in $d$--dimensional Euclidean
space is $\partial$.

\vskip.2cm

\medskip
\noindent{\bf\Theorem (th-ng)} 
{\it Consider  an initial datum satisfying \equ(1.5).
 Then, the sequence of probability measures ${\bf Q}^{{\g}}$ 
is tight and all its limit points ${\bf Q}$
are concentrated  on absolutely continuous 
paths whose densities $\rho \in C^0([0,T]; {\cal M}_1(\TT^d))
\cap L^2\left([0,T],H_1(\TT^d)\right)$ 
are weak solutions of the equation 
$$
\left\{
\eqalign{
& \partial_t \rho  \; =\; \partial\left\{
D(\rho) \left[ \partial \rho -\b\chi(\rho)
      \big(\partial J * \rho \big)\right]\right\}\, ,\cr 
& \rho (0 ,\cdot) \; =\; \rho_0 (\cdot).\, 
}
\right. 
\Eq(1.12)
$$
Moreover if the diffusion coefficient $D$ is 
Lipschitz continuous, then 
${\bf Q}^\g$ converges and the limit point $\bf Q$
is the unique weak solution of \equ(1.12).
     }
\bigskip 

Throughout the text $\Phi$, $J$, $\kappa$ and $\beta$ 
will be considered fixed. We will be often interested
in getting estimates which are uniform on the configuration
$\eta$ or on the history of the process $\{\eta_t\}_{t \ge 0}$.
We therefore introduce the notation $o_u(\cdot)$ and $O_u(\cdot)$
in the standard sense of $o(\cdot)$ and $O(\cdot)$ but uniformly
with respect to the configuration or to the history of the process.

\vskip.5cm
\numsec=3
\numfor=1
\centerline{\bf 3. The fluctuation--dissipation equation.}
\vskip 0.4 cm

In this section we recall the fundamental result
proven in [VY]:
the approximate decomposition of the current in a gradient term
and a fluctuation term.

Let us recall the definition of current  in the general
$J$--dependent case. The current is defined
for every $x$, every unit vector $e$ and every $\eta$ as
$$
w^\g_{x,x+e}(\eta)= c^\g_{x,x+e} (\eta)
[\eta (x)- \eta(x+e)].
\Eq(current1)
$$
It has the property that
$$
{\cal L}_\g \eta (x)=
- \sum_{j=1}^d
\left[
w^\g _{x,x+e_j} (\eta)-
w^\g _{x-e_j,x} (\eta) \right].
\Eq(current2)
$$
The analogous definitions in the case of the unperturbed
dynamics generated by  ${\cal L}^0$
are just obtained by setting $\g =0$.

We follow very closely [VY].
For $\zeta \in \Omega_{\Z^d}$ or $\zeta : \Z^d \longrightarrow {\bf R}$ 
and $\ell \in \IR^+$
we set
$$
{\rm Av}_{\ell}
\zeta (x)=
{1 \over (2\ell +1)^d}
\sum_{y: \vert y-x \vert \le \ell} \zeta (y).
\Eq(defAv)
$$
We also set $\Lambda^\prime_\ell=\{x\in \Z^d: \vert x \vert \le \ell\}$. 
 We define 
the $\sigma$--algebra ${\cal F}_{x,s}$
generated by $\{ {\rm Av}_s \eta (x) \} \cup
\{\eta (y): \vert y-x \vert >s\}
$ and 
the space 
$\cal G$ of local functions $h:\Omega_{\IZ^d} \rightarrow \IR $
 with the property
$\IE^\mu [h \vert {\cal F}_s] =0$ for some $s$.
To any $h \in \cal G$ and 
any $\eta \in \Omega_{\IZ^d}$ we associate an element (still
denoted by $h$)
of $\Omega_{\IZ^d}$ by setting $h(x)=\tau_x h (\eta)$.
We call $\Omega_{\cal G}$ the subset of $\Omega_{\IZ^d}$ obtained 
from $\cal G$  with this procedure.
We then introduce
the finite volume variance
$$
V_\ell (h,\rho, \xi)=
\ell^d 
\langle {\rm Av}_{\ell _1} h , (-{\cal L}^0_{\L^\prime_\ell})^{-1}
{\rm Av}_{\ell _1} h \rangle_{\mu_{\L^\prime_\ell},\rho, \xi},
\Eq(fvvar)
$$
where $\ell_1=\ell-\sqrt{\ell}$, 
${\cal L}^0_\Lambda =\sum_{b \in \Lambda^*} {\cal L}^0_b$
for any finite set $\Lambda\subset \Z^d$
 and  
$\mu_{\L^\prime_\ell,\rho, \xi}$
is the canonical Gibbs measure with interaction $F$
on $\Lambda^\prime_\ell$ with boundary condition
$\xi\in \Omega _{\IZ^d}$ 
and density $\rho\in[0,1]$. 
Two observations are in order for canonical measures. The first is that
we view $\mu_{\L^\prime_\ell,\rho, \xi}$ as an element of
 the probability measures over $\Omega_{\IZ^d}$:
the extension is made   
 in the natural way obtaining a measure 
concentrated on  $\{\eta:\eta (y)=\xi(y)$ for every $y
\in {\L^\prime_\ell} ^c \}$.
Second, if $\rho$ is not an integer multiple
of $1/\vert \L^\prime_\ell\vert$, 
the density of the canonical Gibbs measure
is taken to be  $\max\{k/\vert \L^\prime_\ell\vert: k\in \IZ, 
k\le \rho \vert \L^\prime_\ell \vert\}$.
 We observe then that
 if $x \in \L^\prime_{\ell_1}$,
$h(x)$ depends only on $\eta (y)$, $y \in \L^\prime_\ell$, 
for sufficiently large $\ell$. Moreover $V_\ell
(h,\rho, \xi)$ is an ${\cal F}_\ell$--measurable function
of $\xi \in \Omega$. Given $\mu^\rho$, the infinite volume Gibbs 
measure with density $\rho$, we define 
$V: \Omega_{\cal G} \times [0,1] \rightarrow \IR ^+$ by
$$
V(h, \rho)=\limsup_{\ell \rightarrow \infty}\int \left(
V_\ell (h,\rho, \xi) {\rm d} \mu^\rho (\xi)
\right).
\Eq(preinfvolvar)
$$
Finally we extend this definition to every local function
$h$ by setting  $h_k(x)=h(x)-
\IE^{\mu_{\L^\prime_\ell, \rho,\xi}}[h(x) \vert {\cal F}_{x,k}]$ and taking
the limit
$$
\limsup_{k \rightarrow \infty}
V(h_k, \rho).
\Eq(infvolvar)
$$
Below we use the notation
$$
\nabla _{e_j}  \eta  (x)=
\eta(x+e_j)-\eta (x).
\Eq(notnabeta)
$$

We are now ready to state the {\sl Fluctuation--Dissipation}
Theorem [VY, Theorem 3.4].

\goodbreak \vskip.5cm
\noindent{\bf \Theorem (VYng)} 
{\it 
The symmetric, in fact diagonal, density
dependent matrix $D$ defined in \equ(varform)
coincides with the following Green--Kubo formula:
$$
\chi (\rho) D_{i,j}(\rho)=\delta_{i,j}
{1\over 2} \IE^{\mu^\rho}\left[ c^0_{(0,e_j)}
\right]
-\int_0^\infty \sum_x\IE^{\mu^\rho}
\left[ w^0_{0,e_i} \exp\left(
{\cal L}^0 t \right)
\tau_xw^0_{0,e_j}\right]  {\rm d}t,
\Eq(GKform)
$$ 
where
$\chi$  is the compressibility defined in
\equ(compressibility) and $\delta_{i,j} $ is the Kronecker delta.
Then
for any  
$\a \in \IR^d$
$$
\inf_{{ h}\in {\cal G}^d}
V\left(\sum_{j=1}^d \a_j
\left[
w_{0,e_j}^0 (\eta)
+(D (\rho)\nabla\eta (0))_j
+{\cal L}^0 h_j (\eta)\right],\rho \right)=0.
\Eq(decomp)
$$
Moreover, for any $\delta>0$ there exists 
${ g}^\d : [0,1] \times \Omega_{\IZ^d}
\rightarrow \IR ^d$, 
${ g}^\d (\rho, \cdot) \in {\cal G}^d$ for every $\rho$ and
${g}^\d (\cdot, \eta)$ is smooth for every $\eta$,
such that  
$$
\sup_{\rho \in [0,1]} 
V\left(\sum_j\a_j
\left[w_{0,e_j}^0 (\eta)
+(D (\rho)\nabla\eta (0) )_{j}+
{\cal L}^0 g^\d_j \right],\rho\right)\le\d.
\Eq(approcdec)
$$
}

\vskip 0.3 cm
We observe that, by polarization,
from the bilinear functional $V(\cdot, \rho)$
we can define a scalar product (covariance)
that will be denoted by $\langle \langle f, g \rangle \rangle
(\rho)$, defined for  $f$ and $g$ local function
on $\{0,1\}^{\IZ^d}$.
This covariance is carefully analyzed in Section 8 of
[VY]. We collect here two properties that will be crucial for us.
First, formula (8.7) in [VY] tells us that
$$
\langle \langle
w^0_{0,e_i}, w^0_{0,e_j}
\rangle \rangle
(\rho)=
{1 \over 2}
\delta_{i,j}
{\bf E} ^{\mu ^\rho}
\left[
c_{0,e_j} (\eta) (\nabla _{e_j} \eta (0))^2 \right].
\Eq(property1)
$$
Moreover 
formula (8.13) in [VY]) tells us that
$$
\langle \langle
w^0_{0,e_i}, \nabla_{e_j} \eta (0)
\rangle \rangle (\rho)=
\delta_{i,j}\chi (\rho).
\Eq(property2)
$$

\vskip.5cm
\numsec=4
\numfor=1
\centerline{\bf 4. Proof of Theorem \equ(th-ng)}
\vskip 0.4 cm

{\it 4.1. Preliminary lemmas.}
We will repeatedly need the expansion (in powers of $\gamma$)
of the jump rates: we take advantage of the fact
that an exchange of two particles changes the 
long--range energy of $O_u(\g)$. We will use the following notation
for discrete convolution
$$
(f \star \eta) (x)= \g^d \sum_{z \in \Lambda_\g} 
f\left( \g (x-z) \right) 
\eta (z), 
\Eq(dconv)
$$
where $f$ is a function from $\TT ^d$ to $\IR$ and
$x\in \Lambda_\g$.

\vskip.2cm
\noindent{\bf \Lemma (Taylor)}
{\it For every $\Phi$ and $J$, there exists $C$ such that
for every
$b \in{\Lambda_\g}^*$, every  $\eta\in\Omega_{\Lambda_\g}$ 
and every 
$\gamma \in (0,1)$
$$
\left\vert
\Delta_b \left( \sum_{x,y} J_\g (x,y) 
\eta (x) \eta (y) \right)\right\vert 
\le C \g,
\Eq(preexp) 
$$
and 
$$
\left\vert 
c^\g_{b}(\eta)-c^0_{b}(\eta)
-
{\g\b } \Phi^{\prime}
\left( \kappa \Delta_b H(\eta) \right) 
\left[ \eta(x+e)-\eta(x)\right]
 e \cdot  \left(
\partial J \star \eta\right) (x)
\right\vert
\le C\gamma^{2},
\Eq(expansion1)
$$
where $b=(x,x+e)$.
}

\vskip 0.2 cm

\noindent
{\it Proof.}
The proof follows immediately from the expansion 
in  Taylor series of $\Phi$ \equ(detbalalt).
\qed

\vskip 0.2 cm

A first application of 
Lemma \equ(Taylor) is in proving that the concept of superexponentially
small event coincides for perturbed and unperturbed processes.
As usual, $\beta$ and $\kappa$ are fixed, but recall that
 we  denote by ${\bf P}^{\g,0}$ 
the law of the process in the box $\L_\g$ with $\beta=0$,
that is the reference process in the periodic box.

\vskip 0.2 cm
\noindent{\bf \Lemma (entropyctrl)}
{\it
Let $T$ be a fixed positive number, $\Lambda=\Lambda_\g$ and let 
${\cal B}_{\g,{\bf q}}$ be a bounded functional of the process
$\{ \eta_s \}_{s\in [0,T\g^{-2}]}$ which depends on
 $\gamma>0$ and on a vector parameter 
$\bf q$. Furthermore let $\mu=\mu_\g^{0,
\kappa,\lambda}$ (recall \equ(fullmeasure))
 and let $\mu^\prime$
be any probability measure on $\Omega_{\Lambda_\g}$.
If for every $p \in \IR$
$$
\lim_{{\bf q} } \limsup_{\g \rightarrow 0}
\g^d \log
{\bf E}^{\g,0}_{\mu} \left( \exp\left(p \gamma^{-d} {\cal B}_{\g,{\bf q}}
\right)
\right)=0,
\Eq(assumesuperexp)
$$
then the same is true with ${\bf E}^{\g,0}_{\mu }$ replaced by
${\bf E}_{\mu ^\prime}^\gamma$.
Analogously if
$\{E_{\gamma,{\bf q}}\}_{\gamma, {\bf q}}$ is a family of
${\cal F}_{T\g^{-2}}$ measurable events which are superexponentially
small
under ${\bf P}^{\g,0}_\mu$, that is
$$
\lim_{{\bf q} } \limsup_{\g \rightarrow 0}
\g^d \log
{\bf P}^{\g,0}_{\mu} \left( E_{\gamma,{\bf q}} \right)=-\infty,
\Eq(assumesuperexp')
$$
then they are superexponentially small 
also under ${\bf P}_{\mu^\prime}^\gamma$.  
}

\vskip 0.2 cm
\noindent
{\it Proof.}
First we observe that,  since for some $c=c(F,\kappa, 
\lambda)\in \R$ we have that
$
\sup_{ \mu^\prime, \eta}
({\rm d}\mu^\prime / {\rm d} \mu ) (\eta)
\le e^{c \vert \Lambda_\gamma \vert}
$,
it is sufficient to prove the statements in the
case $\mu^\prime=\mu$. We therefore omit
the subscript $\mu$ in this proof.

For $b=(x,y)$ and $t\ge 0$, let $N_b(t)$ denote the number of jumps 
between $x$ and $y$ in the time span $[0,t]$.
The Radon--Nikodym derivative of ${\bf P}^\g$ with respect to ${\bf P}^{\g,0}$,
both restricted to ${\cal F}_{t\g^{-2}}$ will be denoted by $M_t$ and
it is given by 
$$
M_t=
\exp \left(
-\int_0^{t\g ^{-2}} \sum_b\left[c_b^\gamma (\eta_s)-
c_b^0 (\eta_s)
\right]{\rm d}s+
\int_0^{t\g ^{-2}} \sum_b \log\left(
{c_b^\gamma (\eta_{s^-}) \over
c_b^0 (\eta_{s^-}) }
\right){\rm d} N_b (s)
\right),
\Eq(RNismart)
$$
c.f. [Prop. 2.6, App. 1, KL].
With respect to the measure ${\bf P}^{\g,0}$ and the filtration 
$\{ {\cal F}_t \}_{t \ge 0}$, 
the process $\{M_t\}_{t \ge 0}$ is 
a martingale. Therefore, for $p >1$,  $\{ M_t^p \}_{t\ge 0}$
is a submartingale.
If we define
$$
A_t=\phantom{moveleftmoveleftmoveleftmoveleft}
\Eq(compensator)
$$
$$
-\int_0^{t\g ^{-2}} p  M_s^p \sum_b 
\left[c_b^\gamma (\eta_s)-
c_b^0 (\eta_s)
\right]{\rm d}s+
 \int_0^{t\g ^{-2}}   M_s^p \sum_b 
c_b^0 (\eta_{s}) \left[
e^ {
p \log\left(
{c_b^\gamma (\eta_{s}) \over
c_b^0 (\eta_{s}) } 
\right) 
}-1
\right]
{\rm d} s,
\Eq(comp3)
$$
we have that ${\tilde M}_t=M_t^p-A_t$
is a martingale with ${\tilde M}_0=1$.
We now expand the  expression in the right--hand side
of \equ(comp3) by taking advantage of the fact that 
$(c_b^\gamma (\eta_{s}) /
c_b^0 (\eta_{s}) ) -1=O_u(\gamma)$. Precisely, by Lemma \equ(Taylor),
$$
\exp \left(
p \log\left(
{c_b^\gamma (\eta_{s}) \over
c_b^0 (\eta_{s}) } 
\right)
\right)-1=
p \left( {c_b^\gamma (\eta_{s}) \over
c_b^0 (\eta_{s}) } -1 \right)
+{p(p-1)\over 2}
\left( {c_b^\gamma (\eta_{s}) \over
c_b^0 (\eta_{s}) } -1 \right)^2 +O_u(\gamma^3),
\Eq(comp4)
$$
where we have assumed $p$ bounded, say $p \le 2$, for the last term.
Therefore we obtain that there exists $C$ such that
$$
\eqalign{
A_t
&
=
\int_0^{t\g ^{-2}} M_s^p \sum_b\left[c_b^0 (\eta _s) {p(p-1)\over 2}
\left( {c_b^\gamma (\eta_s) \over c_b ^0 (\eta_s)}
-1
\right)^2 +O_u(\gamma ^3)
\right]{\rm d}s
\cr 
&\le
C( p(p-1) \gamma^{-d+2}+
\gamma^{-d+3})
\int_0^t M_s^p {\rm d}s,
}
\Eq(comp5)
$$
where in the last step 
we have used the fact that $c_b^0(\eta)$ is uniformly bounded
 as well as  the
positivity of $M_t$.
By taking the expectation  of this last expression,
recalling that ${\bf E}^{\g,0}[{\tilde M}_t]=1$,
by Gronwall's Lemma we obtain
$$
{\bf E}^{\g,0}\left[M_t^p\right] \le
\exp\left( Ct \left(p(p-1) \gamma^{-d}+
\gamma^{-d+1} 
\right)
\right).
\Eq(comp6)
$$
This suffices for our purposes since,
by applying H\"older inequality, we obtain
that for every $p>1$ and $q=p/(p-1)$
$$
\eqalign{
\limsup_{{\bf q} } 
\limsup_{\g \rightarrow 0}
\g^d &
\log
{\bf E}^\g \left( \exp(p \gamma^{-d} {\cal B}_{\g,{\bf q}}
\right) \cr
&
\le 
\limsup_{\g \rightarrow 0}
{\g^d\over p} \log
{\bf E}^{\g,0} \left( M_T^p \right)
+
\lim_{{\bf q} } \limsup_{\g \rightarrow 0}
{\g^d \over q} \log
{\bf E}^{\g,0} \left( \exp(q \gamma^{-d} {\cal B}_{\g,{\bf q}}
\right)
\cr
&
\le Ct(p-1).
}
\Eq(compappl)
$$
By letting $p\searrow 1$ the first statement
is proven. The proof of the second statement runs in the same way.
\qed

\vskip 0.2 cm

If $h$ is a local function, we define $\tilde h (\rho)=
\IE^{\mu^\rho}[h]$ for $\rho \in [0,1]$.
Here is the first application of Lemma \equ(entropyctrl).

\vskip.2cm

\goodbreak
\noindent{\bf \Lemma(replacement) }({\it Replacement Lemma}).
{\it Let $h$ be a local function and
$$
{\cal B}_{b\g^{-1}}(\eta)=\left\vert
{1\over (2b\g^{-1}+1)^d}\sum_{|y|\le b\g^{-1}}
\left[
h(\tau_y \eta )-\tilde h\left({\rm Av}_{[b\g^{-1}]}
{\eta} (0) \right)\right]
\right\vert,
\Eq(Bbgamma)
$$
for $b>0$.
Then, for any
$\delta>0$
$$
\limsup_{b\to 0}\lim_{\e\to 0} 
{\bf P}^{\g}\left[
\int_0^T \g ^d \sum_x {\cal B}_{b\g^{-1}} \left( \tau_x \eta _{\gamma^{-2}t}
 \right)
{\rm d} t
\ge
\delta \right]=0.
\Eq(3.2)
$$
}

\vskip 0.2 cm

\noindent{\it Proof.}
This goes through the by now classical 
one block and two blocks estimates. These can
be found in [VY] (Lemma~5.2 and Theorem~6.2)
for the unperturbed process and these estimates
are superexponential. The extension is therefore 
just Lemma \equ(entropyctrl).
\qed

\vskip 0.3 cm

We conclude this subsection with a
computation that is very relevant for us
to identify the limit equation.

\vskip 0.2 cm
\goodbreak
\noindent{\bf \Lemma(algebra) }.
{\it
For any bounded local function $f:\Omega_{\IZ^d} \rightarrow 
{\IR}$ we have that
$$
\IE^{\mu^\rho}\left[ \Phi^\prime (\kappa \Delta_{0,e} H)
\nabla_e \eta(0) 
\Delta_{0,e} f (\eta)
\right]=-{1\over 2}
\IE^{\mu^\rho}\left[ \Phi (\kappa \Delta_{0,e} H)
\nabla_e \eta (0) 
\Delta_{0,e} f (\eta)
\right].
\Eq(primeeq)
$$ 
}

\vskip 0.2 cm
\noindent
{\it Proof.}
By differentiating
both sides of \equ(detbalalt),
we reduce \equ(primeeq) to proving that
$$
E^{\mu^\rho}\left[
\exp\left( -\kappa\Delta_{0,e} H /2\right)
\phi^\prime (\kappa\Delta_{0,e} H) 
\left( \eta (e) -\eta(0)\right) 
\Delta_{0,e} f (\eta)
\right]
=0.
\Eq(step1preq)
$$
Let us now approximate $\mu^\rho$
with a sequence of finite volume grand--canonical Gibbs measures
on $\Omega_{\Lambda_\g}$. The result follows 
because 
$$
\eqalign{
&=
\sum_{\eta\in \Omega_{\Lambda_\g}}
e^{-\kappa \Delta_{0,e} H (\eta) /2}
\phi ^\prime ( \kappa \Delta_{0,e} H (\eta) ) 
\left( \eta (e) -\eta(0)\right) 
\Delta_{0,e} f (\eta)
e^{-\kappa H (\eta)}
e^{\lambda\sum_x \eta (x)}
\cr
&=
\sum_{\eta\in \Omega_{\Lambda_\g}}
\left( \eta (e) -\eta(0)\right) 
\Delta_{0,e} f (\eta)
e^{-\kappa [H (\eta)+ H (\eta^{0,e})]/2}
\phi ^\prime ( \kappa \Delta_{0,e} H (\eta) )
e^{\lambda\sum_x \eta (x)}
\cr
&=
\sum_{\eta\in \Omega_{\Lambda_\g}}
\left( \eta (e) -\eta(0)\right) 
\Delta_{0,e} f (\eta)
e^{-\kappa [H (\eta^{0,e})+ H (\eta)]/2}
\phi ^\prime ( -\kappa \Delta_{0,e} H (\eta) )
e^{\lambda\sum_x \eta (x)},
}
\Eq(useodd)
$$
where in the last step we used the fact that
$( \eta (e) -\eta(0)) 
\Delta_{0,e} f (\eta)$ is invariant under the transformation
$\eta \longrightarrow \eta^{0,e}$.
Recall now that $\phi^\prime$ is odd and 
the proof of 
\equ(primeeq) is complete.
\qed

\vskip 0.3 cm

{\bf 4.2. Tightness and energy estimate.}

\vskip 0.2 cm
We go now to the set up of Theorem \equ(th-ng).
Recall that ${\bf Q}^\g$
is the law of the empirical process $\{ \nu^\gamma (t)\}_{t \in [0,T]}$,
cf. \equ(empfield), on the Skorohod space
$D([0,T],{\cal M})$.
\vskip 0.2 cm

\noindent{\bf \Lemma (tightness)}
{\it
The sequence $\{{\bf Q}^{{\g}}\}_{\g>0}$ 
is tight and every limit point ${\bf Q}$
is concentrated  on absolutely continuous 
paths whose densities belong to

\noindent
$C^0([0,T];{\cal M}_1 (\TT ^d))\cap
L^2 ([0,T],H _1 (\TT^d))$.
} 

\vskip 0.2 cm
{\it Proof.} This is standard. One way to prove tightness
in $D([0,T],{\cal M})$ is to repeat
the proof in [\S 4, VY]: Lemma 4.1 in [VY] depends only on the
uniform boundedness of the jump rates and Lemma 4.2, still
in [VY], is easily upgraded to our situation via
Lemma \equ(entropyctrl). The limit points actually lie
 in a smaller space:
by the exclusion rule it is immediate
to verify that the limit is in ${\cal M}_1$ for every $t$ and,
since every jump produces a discontinuity 
$O_u(\g ^d)$,  we can  substitute
 $D$ with $C^0$.
The {\sl energy estimate}, that is the existence of a constant
 $C$ such that
any limit point ${\bf Q}$ satisfies
$$
{\IE}^{\bf Q}\left[
\int_0^T \int_{\TT^d}(\nabla \rho (s,r))
{\rm d}r{\rm ds}
\right]< C,
\Eq(energyestimate)
$$
requires a more sophisticated  argument.
Once again most of the work has been done in [VY], Section 5:
it is sufficient to upgrade (5.25) and (5.28)--(5.30)
in [VY] to our situation. 
One way would be to get a volume order estimate on
the relative entropy of ${\bf P}^\g$ with
respect to ${\bf P}^0$, both measures restricted
to $[0, T\g^{-2}]$, but we choose to stick to
$L^p$ estimates on the Radon--Nicodym derivative
$M_T$
of the process.
We will be as close as possible to 
the notations of Lemma \equ(entropyctrl) and, like in its proof,
we will omit the dependence on the initial condition:
the only difference is that we will not need any extra
parameter $\bf q$ and we will therefore omit 
the subscript $\bf q$ from the notation.
We choose, with $G\in C^1(\R^+,\TT^d;\R)$ and $C_1>0$, 
$$
{\cal B}_{\g,{\bf q}}={\cal B}_\g=
\g^{d-1} \int_0^{T}
\sum_x G(t, x\g) \nabla_e \eta_{t\g^{-2}} (x)
\dd t-{C_1 \over 2} \g^d
\int_0^T \sum_x
\left\vert G(t,x/N) \right\vert^2 \dd t.
\Eq(chfee)
$$
In the proof of Lemma 3.2 in [VY] it is shown that for some
$C_1$
\equ(assumesuperexp) holds for $p=1$, that is 
$$
\lim_{\g \rightarrow 0}
\g^d \log {\bf E}^{\g,0}
 \left[
\exp \left( {\g^{-d}}
{\cal B}_{\g} \right) \right]=0.
\Eq(abcdef)
$$
Using \equ(abcdef) we will show that
there 
exists $C_2$ such that
$$
\limsup_{\g \rightarrow 0} 
{\bf E}^\g \left[
{\cal B}_{\g} \right] \le C_2.
\Eq(entctrlch)
$$
We refer the reader to [Section 5,VY] for a proof 
of the fact that \equ(entctrlch), with the choice
\equ(chfee), implies \equ(energyestimate) with
$C=\max(C_1,C_2)$. 
Here we prove \equ(entctrlch); it 
is just an application of Jensen's
inequality and Cauchy--Schwarz:
$$
\eqalign{
{\bf E}^\g
\left[
{\cal B}_{\g} \right] &\le 
2\g^d \log {\bf E}^{\g,0} \left[M_T
\exp \left( {\g^{-d}\over 2}
{\cal B}_{\g} \right) \right]
\cr
&
\le 
 \g^d \log {\bf E}^{\g,0} \left[ M_T ^2 \right]
+\g^d \log {\bf E}^{\g,0}
 \left[
\exp \left( {\g^{-d}}
{\cal B}_{\g} \right) \right],
}
\Eq(JCSappl)
$$
and, recalling \equ(abcdef), \equ(entctrlch) follows from 
\equ(compappl) with $p=2$.
\qed

\vskip 0.2 cm
{\bf 4.3. The (partial) identification of the limit.}
By the definition of the process, for every $G\in C^1 (\TT^d ; \IR)$
 we can write
$$
\eqalign{
\g^d \sum_x G(\g x) \eta _{t\g^{-2}} (x)-
&
\g^d \sum_x G(\g x) \eta _0 (x)=
\cr
&\g^{d} \int_0^{t\g^{-2}}
\sum_{x,e} \left[G(\g (x+e))-G(\g x) \right] w^\g_{x,x+e}
(\eta_s) \dd s + M_\g^G (t)
}
\Eq(Mart1)
$$
in which $M_\g^G=\{ M_\g^G (t)\}_{r \in \IR^+}$  is a 
${\bf P}^\g$--martingale
with respect to the filtration associated to $\{ \eta_t \}_{t\in
\IR^+}$. The quadratic variation of $M_\g^G$ is easily
computed and estimated: if we set $X_\g (\eta)=
\g^d \sum_x G(\g x) \eta (x)$ it is immediate to see that
there exists a constant $c=c(d,G)$ such that for every $\g >0$
$$
\langle M_\g^G, M_\g^G  \rangle (t) =
\int_0^{t\g^{-2}} \left(
{\cal L}_\g X_\g ^2
-2 X_\g {\cal L}_\g X_\g \right) (\eta_s)
\dd s \le c \gamma^d,
\Eq(boundqv)
$$
and thus, by Doob's inequality, for every $T>0$ and every
$\delta>0$
$$
\lim_{\g \rightarrow 0} {\bf P}^\g \left(
\sup_{t\in [0,T]} 
\left\vert M_\g ^G (t) \right\vert
>\delta 
\right)=0.
\Eq(Doobs1)
$$

We now split the current: 
$$
w^\g=[w^0]+
[w^\g-w^0].
\Eq(split)
$$
The non gradient difficulties  come 
from the first term. 
Below we use the convention that $\sum_e$ is the sum over
the unit vectors $\{e_j\}_{j=1,2,\ldots, d}$.

We first have a fast look at 
the {\sl easy} second term. 
By Lemma \equ(Taylor), for every $x$ and every unit vector $e$
$$
[w^\g-w^0]_{x,x+e}(\eta)=
-{\beta}\gamma
\Phi^\prime \left( \Delta_ {x,x+e} H(\eta)\right)
[\nabla_e \eta]^2(x)
e\cdot (\partial J \star \eta)(x)+ O_u(\g ^2).
\Eq(wgminusw0)
$$
If we define
$$
\eqalign{
{\cal R}_{b,\g}\left( \left\{ \eta_s \right\}_{s \ge 0}\right)
&=
\Bigg\vert
\int_0 ^{\gamma^{-2}T}
\gamma^d \sum_{x,e }
\left[ G(\g (x+e))-G(\g x)  \right]\left[
w^\g_{x,x+e}(\eta_t)-w^0_{x,x+e} (\eta_t)
\right] {\rm d}t 
\cr
 +\beta
\int_0^{\gamma^{-2}T} 
&    
\gamma^{d+2}
\sum_{x,e} \left( \partial_e G \right) (\g x) 
\left( \partial _e J \star \eta_t \right) (x)
{1 \over (2b \gamma^{-1}+1)^d}
\sum_{\vert y \vert \le b \g^{-1}}
h_e \left( \tau_{x+y} \eta _t \right) 
{\rm d} t
\Bigg\vert,}
\Eq(Rbgamma)
$$
where $h_e (\eta)= \Phi^\prime (\Delta_{0,e} H(\eta))
(\eta (e)-\eta(0))^2$, and therefore
$$
\tilde h _e (\rho)=
\IE^{\mu^\rho}
\left[
\Phi ^\prime \left( \kappa \Delta_{0,e} H(\eta) \right)
(\eta (e)-\eta(0))^2\right].
\Eq(tildehp)
$$
Note that, by applying Lemma \equ(algebra), with $f\equiv 1$,
we can immediately rewrite
$$
\tilde h _e (\rho)=-{ 1\over 2}
\IE^{\mu^\rho}
\left[
c_{0,e}^0 (\eta)
(\eta (e)-\eta(0))^2\right].
\Eq(tildehpr)
$$
By the smoothness of $G$ and $J$ it is immediate to
obtain that
$$
\lim_{b \rightarrow 0} \limsup_{\g \rightarrow 0}
\sup_{\left\{ \eta_s \right\}_{s \ge 0}}
{\cal R}_{b,\g}\left( \left\{ \eta_s \right\}_{s \ge 0}\right)
=0.
\Eq(Rbgammagoesto0)
$$
This, together with Lemma \equ(replacement) applied to the spatial average of 
$h$, immediately implies that
$$
\eqalign{
\Bigg\vert 
\int_0 ^{\gamma^{-2}T}
\gamma^d \sum_{x,e }
\left[ G(\g (x+e))-G(\g x)  \right]
&
\left[
w^\g_{x,x+e}(\eta_t)-w^0_{x,x+e} (\eta_t)
\right] {\rm d}t -
\cr
\int_0^T \int_{\TT^d}  {\partial } G (r)
&
( {\partial } J \star \nu^\g )(t,r)
\tilde h 
\left( {\bf 1}_{(-b,b)^d} \star \nu ^\g ) (t,r) \right)
{\rm  d}r {\rm d}t \Bigg\vert}
\Eq(goesto0inP)
$$
tends to zero in ${\bf P}^\g$--probability as $\g \rightarrow 0$
and then $b\rightarrow 0$.

Now we turn to the non gradient term.
Given $a$, $b$, $\delta$ and $\delta_1>0$, we define
$$
\eqalign{
{\cal B}_{a,b,\delta }=
\Bigg\{ 
&
\sup_{t \in [0,T]} \Bigg\vert
\g ^{d+1} \int_0^{\g^{-2}t} \sum_x
\sum_e \partial_e  G (\g x) w^0_{x,x+e} (\eta_t) {\rm d}t+
\cr
 \gamma ^{d+1} \int _0^{\g^{-2}T} (2b)^{-1}
\sum_{x,e,e^\prime} 
&
\left[ \eta_s (x+b\gamma^{-1} e)-
\eta_s (x-b\gamma^{-1} e) \right]
D_{e, e^\prime}
\left( 
{\rm Av}_{[a \g^{-1}]}
{\eta}_s (x)
\right) \partial_e G( \gamma x) +
\cr
& \gamma ^{d+1} \int _0^{\g^{-2}T}
\sum_{x,e} \partial_e G (\g x)
{\cal L}^0 g^\delta_e \left(
\tau_x \eta _t \right)
\Bigg\vert \ge \delta_1
\Bigg\}.
}
\Eq(Babdd1)
$$ 
The following  result follows from Theorem 3.3 in
[VY] and Lemma \equ(entropyctrl), with the only observation that 
in the statement in [VY] the term ${\cal L}^0 g^\delta$
does not appear: in their case it is irrelevant
(see Section 3 of [VY]).

\vskip 0.2 cm

\noindent{\bf \Lemma (Babdd)}
{\it
For every $\delta_1>0$
$$
\lim_{a\rightarrow 0}
\limsup_{b\rightarrow 0} \limsup_{\delta \rightarrow 0}\limsup _{\g \rightarrow 0}
{\bf P}^\g \left(
{\cal B}_{a,b,\delta}\right)=0.
\Eq(Babdgt0)
$$
} 

\vskip 0.2 cm
We now observe that in \equ(Babdd1) we can replace
${\cal L}^0 g^\delta$ with $({\cal L}^0 - {\cal L}_\g) g^\delta$
and Lemma \equ(Babdd) still holds
since
$$
\eqalign{
&\phantom{movemovemov}
\g^{d+1} \int_0^{\g^{-2} T}
\sum_{x,e} \partial_e G (\g x) {\cal L}_\g  g^\delta_e 
\left( \tau_x \eta_t \right)
\cr
=
\g &\left[
\g^d \sum_{x,e} \partial_e G (\g x)
\left(
g^\delta_e \left(\tau_x \eta _{\g^{-2} T}\right)
-g^\delta_e \left(\tau_x \eta _{0}\right) \right)
 \right]
+ \g {\tilde M}_\g (T) + E_\g (T).
}
\Eq(Lgcont0)
$$
where
$\{ \tilde M^\g (t)\}_{t\ge 0}$ with respect to the
natural filtration is a ${\bf P}^\g$--martingale
and $E_\g(T)$ is the error we made by considering $g^\delta$
independent of $\{{\rm Av}_{\ell_\delta}{\eta}(x)\}_x$,
$\ell_\delta$ the range of $g^\delta$.
The first term in the second line of \equ(Lgcont0)
is $O_u(\g)$, the martingale ${\tilde M}_\g$ has a quadratic
variation $O(\g ^d)$ and $\sup_{t \in [0,T]} \vert E_\g (t)\vert$ 
tends to zero in probability
as $\gamma \rightarrow 0$ and $\delta \rightarrow 0$.
The last claim follows from the argument in [VY],
formulas (3.26)--(3.27), and Lemma \equ(entropyctrl).

We are therefore left with evaluating
$$
\gamma
\int_0^{\g^{-2} T}
\gamma ^d \sum_{x,e} \partial _e G (\g x) \left[{\cal L}^0
-{\cal L}_\g \right]
g^\delta_e \left( \tau_x \eta _t\right)
{\rm d }t.
\Eq(wil)
$$
By using Lemma \equ(Taylor) to express 
${\cal L }^0 -{\cal L}_\g$, recalling that $g^\delta _e$
is a local function and
repeating exactly the same steps, i.e. {\sl smearing } and using
the replacement lemma,
as in \equ(Rbgamma)--\equ(goesto0inP),
we obtain that, up to a negligible error term,
 this term can be replaced by
$$
\beta \sum_{e, e^\prime}
\int_0^T
\int_{\TT^d} \partial_eG(r)
\theta^\delta_{e, e^\prime}
\left(
\chi_{[-\delta, \delta]^d} * \nu ^\g (t,r)
\right)
\partial _{e^\prime} J * \nu^\g (t,r) {\rm d}r {\rm d} t,
\Eq(wilsubs)
$$
where 
$$
\theta^\delta_{e,e^\prime} (\rho)=
{\bf E}^{\mu^\rho}
\left[
\sum_z
\left( \eta (z+e^\prime) -\eta(z)\right) 
\Phi^\prime \left( \kappa
\Delta_{z,z+e^\prime} H_0 (\eta) \right)
\left[g_e^\delta (\eta^{z,z+e^\prime})-
g_e^\delta (\eta) \right]\right].
\Eq(h2e)
$$

\vskip 0.2 cm
Let us define $\theta (\rho) =\limsup_{\delta \rightarrow 0} 
\theta ^\delta (\rho)$ for every $\rho \in [0,1]$.
By Lemma \equ(tightness) and collecting the results 
\equ(Mart1), \equ(Doobs1), \equ(Rbgammagoesto0) and 
Lemma \equ(Babdd), under the hypothesis that  
$ \theta^\delta $ converges uniformly 
as $\delta \rightarrow 0$,
we obtain that every limit of the sequence $\{\nu^\g\}_{\g>0}$ 
concentrates on trajectories $\rho$ which solve weakly
the PDE
$$
\left\{
\eqalign{
& \partial_t \rho  \; =\; \sum_{i,j} {\partial}_i
\left\{
D_{i,j}(\rho) 
    { \partial}_j \rho -\b
     \left( \delta_{i,j}
    {\tilde h }_{j} (\rho)+  \theta_{i,j}(\rho)
    \right)
    \left( \partial _j J * \rho \right)
\right\}
\, ,\cr 
& \rho (0 ,\cdot) \; =\; \rho_0 (\cdot).\, 
}
\right. 
\Eq(1.12prime).
$$
We are therefore left with showing that
\vskip 0.3 cm

\noindent{\bf \Lemma (identity)}
{\it The sequence $\theta^\delta$ converges uniformly and
for every $\rho\in [0,1]$, every $i$ and every $j =1,2,
\ldots, d$ we have that
$$
\delta_{i,j}
{\tilde h} _{j} (\rho)+  \theta_{i,j}(\rho)=
\chi (\rho)D_{i,j}(\rho).
\Eq(identityfor)
$$
}
\vskip 0.3 cm

\noindent{\it Proof.}
We divide this algebraic computations in steps

\noindent{\it Step 1}. We start by rewriting
$\theta$ in a more convenient form, using 
Lemma \equ(algebra) with $f=g^\delta_{e_j}$.
We obtain that
$$
\theta _{i,j}^\delta (\rho)=
-{1\over 2}
{\IE}^{\mu^\rho}
\left[\sum_z
c_{z,z+e_j}(\eta) \left( \eta (z+e_j)-\eta (z) \right)
\Delta_{z,z+e_j} g_{e_i}^\delta (\eta)
\right].
\Eq(step1)
$$

\vskip 0.2 cm

\noindent 
{\it Step 2.}
Reversibility and summation by parts on \equ(step1) imply
$$
\theta_{i,j}^\delta (\rho)=
{\IE}  ^{\mu^\rho}
\left[\sum_z
c_{z,z+e_j}(\eta) \left( \eta (z+e_j)-\eta (z) \right)
g_{e_i}^\delta (\tau_z \eta)
\right]=
-
\langle w^0_{0,e_j},g_{e_i}^\delta
\rangle _0 (\rho),
\Eq(step2.1)
$$
where in the last step we used the notation 
$$
\langle f,g \rangle_0 (\rho)=
\sum_x
{\IE}^{\mu^\rho}\left[
f(\eta) g(\tau_x \eta) \right],
\Eq(notfg)
$$
for $f$ and $g$ local functions.
It is not too difficult to show
[Section 8, VY] that
$$
\langle f,g \rangle _0=
-
\langle \langle f,{\cal L}_0g \rangle \rangle,
\Eq(Step2.re)
$$
and we recall that the {\sl covariance} $\langle \langle \cdot,\cdot
\rangle \rangle$ is defined 
right after Theorem~\equ(VYng).
Therefore
$$
\theta^\delta_{i,j}=
\langle \langle
w^0_{0,e_j}, {\cal L}_0 g_{e_i}^\delta
\rangle \rangle.
\Eq(Step2.2)
$$

\vskip 0.2 cm
\noindent{\it Step 3.}
We now use the decomposition induced
by \equ(decomp) to express ${\cal L}_0 g_{e_j}^\delta$
obtaining that $\theta^\delta _{i,j}$ is equal  to
$$
\langle \langle
w^0_{0,e_j}, {\cal L}_0 g_{e_i}^\delta
\rangle \rangle (\rho)
=
\langle \langle
w^0_{0,e_i}, w^0_{0,e_i}
\rangle \rangle
(\rho)
-
\langle \langle
w^0_{0,e_j}, (D\nabla \eta)_{e_i} (0)
\rangle \rangle (\rho)
+R(\rho, \delta),
\Eq(Rrhodel)
$$
where $\lim_{\delta \rightarrow 0} 
\sup_\rho \vert R (\rho, \delta) \vert=0$.

Completing the proof of \equ(identityfor)
 is now just a matter of applying \equ(property1),
note the cancellation with the term $\tilde h$, c.f. \equ(tildehpr),
and \equ(property2)
and the proof is complete.
\qed

\vskip 0.2 cm
Lemma \equ(identity) leads us to the weak formulation
of the PDE \equ(1.12), that is that every limit measure $\bf Q$
is concentrated on trajectories $\rho\in
 C^0([0,T]; {\cal M}_1(\TT^d))
\cap L^2\left([0,T],H_1(\TT^d)\right)$ 
which solve 
$$
\eqalign{
\int_{\TT^d} G(r)\rho(T,r) {\rm d}r
-&
\int_{\TT^d} G(r) \rho(0,r) {\rm d}r=
\cr
&
\int_0^T
\int_{\TT ^d}
\partial G(r)\left(
D(\rho(t,r))\chi(\rho(t,r))
\partial J * \rho - D(\rho) \partial \rho (t,r)
\right){\rm d}r {\rm d}t,}
\Eq(weakform)
$$ 
for every $G \in C^1 (\TT^d)$ and every $T >0$.
Therefore the proof of Theorem
\equ(th-ng) is complete modulo discussing
the uniqueness issue. This is considered in the next subsection.

\vskip 0.2 cm
{\bf 4.4. Uniqueness.}
A proof of uniqueness is available if $D$ is a uniformly Lipschitz 
continuous function. This follows from the control of
$H^{-1}$ norm. In [GL2] this result is proven under an additional
condition on the time derivative, in the sense of distributions,
of $\rho$. This condition can be removed if one replaces 
the kernel of the $H^{-1}$ norm with a smoothened kernel, as 
it is done in [Appendix A, KL]. Note that the weak formulation 
\equ(weakform) is in this set up equivalent to
the formulation that we get by multiplying
both sides of \equ(1.12) by $\tilde G \in C^1(  \IR^+\times
\TT^d; \IR)$ and 
formally integrating by parts.

\vskip.5cm
\numsec=5
\numfor=1
\centerline{\bf 5. Multispecies systems}
\vskip 0.4 cm

The scheme of proof of Section 3 and Section 4 
can be applied to several other systems.
Of particular interest for applications
is the case of systems with several types
of particles. One (apparently) simple system 
in this class is the $n$--color exclusion process:
take a simple exclusion process and distinguish
the particles by painting them with $n$ colors ($n$ is kept
fix).  The hydrodynamics of this system has been 
done in [Q]: here we would like to consider the case
in which the particles interact via long
range potentials that distinguish between colors.

$ $From the technical viewpoint
this case has essentially been already
considered in [QRV], where a Large Deviation principle
for a $n$--color exclusion system is proven as  a step
to prove a process Large Deviation for the simple 
exclusion. The lower bound of the Large Deviations
depends on the standard change of measure argument
which entails studying the $n$--color system 
driven by a weak external force. This is also our case:
the weak driving force is however configuration
dependent, but since it depends on the configuration
only via an empirical average, the changes with 
respect to [QRV] are minimal. We will therefore
simply state the result and make some observations
on the proof.

To simplify the notation and the statement of the result we restrict
our attention to the case of two species (A and B) and to
the case in which A and B interact with each other, but A does not
interact with any A and the same for B particles.

\vskip 0.2 cm
\noindent
{\bf 5.1 The A--B model.}
As in the previous sections $\Lambda_\g$ will denote the
lattice torus with diameter $[\gamma^{-1}]$.
We are looking at a $d$--dimensional system of A and B particles 
evolving via a Kawasaki dynamics with Kac Hamiltonian
$$
H_\gamma (\eta)={1\over 2}
\sum_{x,y\in \Lambda_\g} J_\gamma (x,y)
\eta^A(x) \eta^B(y)
\Eq(HAB)
$$
where as usual $\gamma >0$ and 
$\eta^A$ and $\eta^B$ are elements of $\{0,1\}^{\Lambda_\g}$
with the hard core restriction that there can be at most one particle per site
$$
\eta^A +\eta^B\le 1 .
\Eq(costr)
$$
$J_\g(x,y)=\g^dJ(\g(x-y))$, where $J$ is a smooth function from the $d$--dimensional unit torus to $\IR$. 
  The particle configuration can be alternatively
described by
$$
\eta \in \{0,A,B\}^{\Lambda_\g},
\Eq(eta)
$$
and we will identify $\eta $ with $(\eta^A ,\eta^B)$.
The dynamics which conserves both $A$ and $B$ particles (or interchanging
with) are specified by $A$ and $B$ particles hopping to nearest neighbor
empty sites at a rate $D c_b^{\g, AB} (\eta)$  but no
direct exchanges between $A$ and $B$ particles are permitted.  This models
the physical situation of polymers in a fluid considered in [WD].  
The generator of the dynamics is
$$
L_\gamma f(\eta) =
\Eq(genera)
$$
$$
D
\sum_{b=(x,y)\in \Lambda_\g^*} c_b^{\g, AB} ( \eta) 
\left[(\eta^A(x)-\eta^B(y))^2+
(\eta^B(x)-\eta^A(y))^2\right]
\left[ f(\eta^{x,y})-f(\eta) \right],
$$
where $D>0$, $f$ is a real valued function
defined in $\{0,A,B\}^{\Lambda_\g}$
and $\eta ^{x,y}$ is $\eta $ with the occupation
of the sites $x$ and $y$ exchanged. As before, the rates 
$c_b^{\g, AB}$ are such that the dynamics satisfies
the detailed balance condition
with respect to the Gibbs measure 
with Hamiltonian \equ(HAB) at inverse temperature
$\beta >0$:
$$
c_{(x,y)}^{\g,AB} (\eta )
=\Phi
\left( \beta 
\left[
H_\gamma (\eta^{x,y} ) - H_\gamma (\eta)
\right]\right),
\Eq(ratesAB)
$$
$\Phi$ as in \equ(detbalalt).
To denote the process we keep the same notation
of the previous sections.

\vskip 0.2 cm
\noindent
{\bf 5.2 The hydrodynamics of the A--B model.}
For the hydrodynamic limit
we look at the empirical measure
$$
\nu^\g_\alpha (t,x)=\g^d
\sum_{y \in \Lambda_\g}
\delta (x-\g y)
\eta^\alpha_{\g^{-2} t } (y),
\Eq(empmeasAB)
$$
for $\alpha\in \{A,B\}$, 
$x\in \TT^d$  and $\eta^\alpha (x,s)$ specifies the presence or absence of an
$\alpha$-particle on 
site $x$ at time $s$.
 
For the initial datum we assume
$\nu^\g_{\alpha} (0,\cdot)$
to be close to $\rho^\alpha(\cdot)$ in the sense of 
\equ(1.5). In this case we have to make further
assumptions on $\rho^\alpha$, precisely that there exists $\delta>0$
such that for every $x$ and every $\alpha$
$$
\delta\le \rho^\alpha (x) \le 1-\delta,
\Eq(awayfromand1)
$$
and that $\rho^\alpha$ is differentiable with bounded derivatives 
$$
\sup_{x, \alpha} \left\vert \partial \rho^\alpha (x) \right\vert
<\infty.
\Eq(bdder)
$$

For the moment let us  set $J=0$.
In [Q] it is shown that
$\nu^{\g}_{\alpha}$ converges {\sl weakly in probability}, that is 
in the sense of \equ(1.5) for every $t\ge 0$, 
 to 
$\rho^\alpha :\IR^+\times \TT^d \rightarrow [0,1]$
and the couple $(\rho^A, \rho^B)$ solves
$$
\partial _t
\left(
\matrix{ \rho^A \cr
\rho^B }
\right)=
\partial\cdot \left[
\left(
\matrix{
{\rho^B \over \rho} D_s +
{\rho^A \over \rho}D & {\rho^A \over \rho}(D-D_s)
\cr
{\rho^B \over \rho}(D-D_s) &
{\rho^A \over \rho} D_s+
{\rho^B \over \rho}D
}
\right)
\partial \left(
\matrix {\rho^A \cr
\rho^B 
}
\right)\right],
\Eq(diffusion)
$$
with initial condition $(\rho^A(0,\cdot), \rho^B(0, \cdot))=
(\rho^A(\cdot), \rho^B(\cdot))$.
In \equ(diffusion) $\rho=\rho^A+\rho^B$ and $D_s=D_s(\rho)$ is
the self diffusion coefficient.
The expression for the diffusion matrix $D$ in \equ(diffusion)
can be derived from elementary considerations
on the microscopic system from which it is derived 
[LS]. We observe that, as expected, the evolution equation for $\rho$ is
simply
$$
\partial_t \rho =D \Delta \rho.
\Eq(heat)
$$
This follows from the observation that if we ignore the distinction between
$A$ and $B$ particles then, in the absence of interactions, we are just
dealing with the one component simple symmetric exclusion process [Sp].  

We can rewrite 
the system \equ(diffusion)
as 
$$
\partial_t \underline{\rho}
=\nabla \cdot\left[ {\bf {\cal D}}
\nabla  \underline{\rho} \right]
=\nabla \cdot\left[ {\bf M}
\nabla  {\delta {\cal F}_0 \over \delta \underline{\rho}} \right]
\Eq(notation)
$$
in which $\underline{\rho}= (\rho^A , \rho^B)^t$,
and 
$$
{\cal F}_0(\rho^A, \rho^B)=
-{1\over \beta }\int s(\rho^A, \rho^B) {\rm d}x,
\Eq(free0)
$$
with
$$
s(\rho^A, \rho^B)=-
\left[ \rho^A \log \rho^A +\rho^B \log \rho^B+
(1-\rho^A-\rho^B) \log (1-\rho^A-\rho^B)\right]
\Eq(free0aux)
$$
and
$$
{\bf M}= -\beta {\cal D} ({\bf Hess}(s))^{-1}.
\Eq(M)
$$
Explicitely
$$
{\bf Hess}(s)^{-1}
= \left(
\matrix{
-\rho^A (1-\rho^A) & \rho^A \rho^B
\cr
\rho^A \rho^B & -\rho^B(1-\rho^B)
}
\right),
\Eq(expl)
$$
and 
$$
{\bf M}=
\beta
\left(
\matrix{
D_s {\rho^A \rho^B \over \rho} +
D {(\rho^A)^2 (1-\rho) \over \rho }
& 
-D_s {\rho^A \rho^B \over \rho}+
D
{\rho^A \rho^B (1-\rho) \over \rho}
\cr
D {\rho^B \rho^A (1-\rho) \over \rho}
-D_s {\rho^A \rho^B \over \rho}
&
D{(\rho^B)^2 (1-\rho) \over \rho}+
D_s {\rho^A \rho^B \over \rho}
}
\right).
\Eq(expl2)
$$

We now claim that,
in the case in which $J$ is not zero,
the limit equation \equ(notation) has to be changed 
by replacing ${\cal F}_0$ with 
$$
{\cal F}={\cal F}_0 
+{1\over 2} \int \int J(x-x^\prime) \rho^A (x) \rho^B(x^\prime)
{\rm d} x {\rm d}x^\prime.
\Eq(free)
$$

\goodbreak \vskip.3cm
\noindent{\bf \Proposition (fromQRV)} 
{\it Set $d\ge 3$.
For every $t\ge 0$ 
the empirical field $(\nu^\g_a (t, \cdot), \nu^\g_B (t,\cdot))$
converges weakly in probability, i.e. in the sense of \equ(1.5), 
to the unique weak solution of
$$
\partial_t \underline{\rho}
=\nabla \cdot\left[ {\bf {\cal D}} \partial \underline{\rho}
+{{\bf M} } \partial  J *\underline{\rho}
\right]. 
\Eq(general1)
$$
}
\vskip .3 cm

The restriction to $d \ge 3$ is due to the fact
that only under this restriction  $D_s$ is known to
be a Lipschitz function and uniqueness of the
weak solution holds.

\vskip1cm
{\bf Acknowledgments}.  R. M. would like to 
thank R. Esposito and E. Presutti for useful discussions.
Research supported by NSF Grant DMR-9813268, AFOSR Grant F49620-98-1-0207,
and DIMACS and its supporting agencies the NSF under contract STC--91--19999
and the NJ Commission on Science and Technology.
G.G. acknowledges the support of MURST (cofin99) and of GNAFA and R. M. the support of 
MURST (cofin98).

\goodbreak
\vskip1truecm
\noindent {\bf References}
\medskip

\item{[A]} A. Asselah and L. Xu, private communication. 

\item{[BL]}
P. Butt\`a, J. L. Lebowitz, {\it Hydrodynamic limit
of Brownian particles interacting with short-- and long--range forces}, 
J.
Statist. Phys. {\bf 94} (1999), no. 3--4, 653--694.

\item{[DG]} P. G.  de Gennes, {\it  Dynamics of fluctuations 
and spinodal decomposition in polymer blends}, J. Chem. Phys. {\bf 72}
 (1980), no. 9,
4756--4763.



\item{[EP]} W.  E and P. Palffy-Muhoray, {\it Phase separation in 
incompressible systems}, Phys. Rev. E (3) {\bf 55} (1997), 
no. 4, R3844--R3846.

\item{[GL1]} G. Giacomin  and J. L. Lebowitz, 
{\it Phase segregation dynamics in particle systems with long range interaction I.
Macroscopic limits},  J. Stat. Phys. {\bf 87}, 37-61. (1997).

\item{[GL2]} G. Giacomin  and J. L. Lebowitz, 
{\it
 Phase segregation dynamics in particle systems with long range interaction II.
Interface motion}, { SIAM J. Appl. Math.} {\bf 58}, 1707--1729 (1998).

\item{[GLP]} G. Giacomin, J. L. Lebowitz and  E. Presutti, 
{\it Deterministic and Stochastic Hydrodynamic Equations 
Arising From Simple Microscopic Model Systems},
{ Stochastic partial differential
equations: six perspectives}, 107--152, Math. Surveys Monogr., 64,
Amer. Math. Soc., Providence, RI, 1999.

\item{[GPV]}  M. Z. Guo, G. C. Papanicolaou and S. R. S. Varadhan,  
{\it Nonlinear diffusion limit for a system with nearest 
neighbor interactions},
{ Comm. Math. Phys.}, {\bf 118}(1988), 31-59.

\item{[KL]} 
C. Kipnis and C. Landim, 
{\it Scaling limits of interacting particle systems},
 Grundlehren der Mathematischen Wissenschaften
[Fundamental Principles of Mathematical Sciences], {\bf 320}, Springer-Verlag,
 Berlin, 1999. 

\item{[Li]}
T. Liggett, {\it Interacting particle systems},
Springer--Verlag (1985).

\item{[LOP]}
J. L. Lebowitz, E. Orlandi, E. Presutti, {\it A
particle model for spinodal decomposition}, J. Statist. Phys. {\bf 63}
 (1991), no.
5--6, 933--974. 

\item{[LS]}
J.L.~Lebowitz and H.~Spohn,
{\it
Comment on: "Onsager reciprocity relations without 
microscopic reversibility" by D. Gabrielli, G. 
Jona--Lasinio and C. Landim},
Phys. Rev.
Lett. {\bf 78} (1997), no. 2, 394--395.

\item{[MM]} R. Marra and M. Mourragui, {\it Phase segregation 
dynamics for the Blume-Capel model with Kac interaction.}
to appear on Stoc. Processes and Appl. (2000).

\item {[Q]} J. Quastel, 
{\it Diffusion of a color in the simple exclusion process},
 Comm. Pure Appl. Math. {\bf XLV},
623--679 (1992).

\item{[QRV]}
J. Quastel, F. Rezakhanlou and S. R. S.
Varadhan, {\it
Large Deviations for the symmetric
simple exclusion process in dimension $d \ge 3$},
 Probab. Theory Related Fields {\bf 113} (1999), no. 1, 1--84.

\item{[R]}
F. Rezakhanlou, {\it Hydrodynamic limit for a
system with finite range interactions}, Comm. Math. Phys. {\bf 129}
 (1990), no.
3, 445--480.

\item {[Sp]} H. Spohn,  {\it Large scale dynamics of interacting particles},
Springer, Berlin, (1991).

\item{[V]} 
S.R.S. Varadhan, {\it Nonlinear diffusion
limit for a system with nearest neighbor interactions II},
in {\sl Asymptotic Problems in Probability Theory~: Stochastic Models
and Diffusion on Fractals}, edited by K. Elworthy and N. Ikeda,
Pitman Research Notes in Mathematics {\bf 283}, Wiley, (1994).

\item{[VY]}
 S. R. S. Varadhan and H.--T. Yau, 
{\it Diffusive Limit of Lattice Gas with Mixing Conditions}.
 Asian J. Math. {\bf 1}, 623--678 (1997).

\bye